\documentclass[aps,twocolumn,prl,superscriptaddress,tightenlines,longbibliography]{revtex4-1}
\usepackage{CJK}
\usepackage{amsfonts}
\usepackage{graphicx}
\usepackage{hyperref}
\usepackage{amsmath}
\usepackage{amssymb}
\usepackage[usenames, dvipsnames]{color}
\usepackage{subfigure}
\usepackage{lipsum}
\usepackage{lineno}

\begin{document}

\title{Non-Markovian Multiphoton Chiral Dynamics with Giant Systems}
\author{Yue Chang}
\email{yuechang7@gmail.com}
\affiliation{Beijing Academy of Quantum Information Sciences, Beijing
100193, China}

\begin{abstract}
Chiral interactions enable directional control of quantum light, providing
new routes for nonreciprocal photon dynamics. While previous studies focused
on single photon regimes, this work explores the non-Markovian multiphoton
regime in a giant nonlinear system nonlocally coupled to a one-dimensional
waveguide, with a tunable phase difference breaking parity symmetry. We show
that single-photon transport remains reciprocal, but multiphoton dynamics
become chiral, generating direction-dependent higher-order correlations and
enabling deterministic creation of multiphoton states with nontrivial
statistics from a single input direction, even under strong dissipation. The
non-Markovian nature allows tuning of transmitted photon statistics via the
coupling-point separation. At a specific phase, internal dynamics become
Markovian while output photons retain non-Markovian features. Under strong
coherent driving, we uncover nonreciprocal dissipative phase transitions and
demonstrate how non-Markovianity shapes the critical response. This
framework offers a general strategy for controlling chiral multiphoton
dynamics in nonlocally coupled waveguide QED systems.
\end{abstract}

\maketitle

\section{Introduction}

The ability to engineer single-photon-level light-matter interactions in
confined geometries is essential for quantum information processing, with
applications ranging from cat-state generation to quantum communication
protocols \cite%
{milonni2013quantum,kimble1998strong,Walls2008,haroche2006exploring}. A
particularly promising direction is chiral quantum optics, where emitters
couple directionally to propagating modes in one-dimensional (1D) media \cite%
{PhysRevLett.70.2269,PhysRevLett.70.2273,PhysRevA.105.023712,lodahl2017chiral,roushan2017chiral,kannan2023demand}%
. Such direction-dependent interactions offer new routes for generating
exotic quantum many-body states \cite{Stannigel_2012,PhysRevLett.113.237203}
and realizing cascaded quantum networks \cite{PhysRevLett.78.3221,Kimble2008}%
. Early realizations of chiral coupling relied on nanophotonic systems,
where polarization-selective waveguides enforced directionality at the
single-emitter level \cite%
{Petersen2014,PhysRevX.5.041036,Scheucher2016,lodahl2017chiral}. In these
systems, transitions between certain emitter states can be induced
exclusively by light traveling in a specific direction.

A further promising route to achieving chiral interactions is through a
waveguide QED architecture coupled to a giant emitter \cite%
{PhysRevA.90.013837,PhysRevA.95.053821,PhysRevLett.120.140404,kannan2020waveguide,PhysRevA.105.023712,PhysRevA.106.033522,PhysRevA.106.013702,PhysRevA.106.063717,PhysRevResearch.3.043226,PhysRevX.13.021039}%
. Unlike \textquotedblleft small\textquotedblright\ emitters, giant emitters
couple to multiple, spatially separated points along a waveguide. Owing to
interference between propagating fields between the coupling points, giant
emitters can exhibit a rich variety of physical phenomena, including a
frequency-dependent Lamb shift \cite{PhysRevA.90.013837}, nonexponential
atomic decay and distinctive single- and two-photon scattering properties
\cite{PhysRevA.95.053821}, topological and retardation effects \cite%
{PhysRevA.106.033522}, and nonreciprocal frequency conversion \cite%
{PhysRevResearch.3.043226}. Experimentally, recent advances in
superconducting platforms have enabled the realization of giant artificial
atoms based on transmon qubits \cite%
{Gu2017,kannan2020waveguide,PhysRevX.13.021039}. It has been shown that
time-modulated interactions allow precise control over the relative phase $%
\phi $ between coupling points \cite{PhysRevX.13.021039}. This nonzero $\phi
$ breaks parity symmetry and enables tunable, chiral light--matter
interactions.

Despite this parity-symmetry breaking, single-photon transport remains
reciprocal due to an underlying parity-time ($\mathcal{PT}$) symmetry \cite%
{2020npjQI...6...32G,PhysRevLett.126.043602,chen2022nonreciprocal,PhysRevA.106.033522,Wang_2022,PhysRevLett.133.063603}%
. That is, the reflection amplitude is identical for left- and
right-incident single photons. In contrast, multi-photon processes break
this reciprocity. In this work, we explore the nonlinear, driven dynamics of
a giant chiral system in the few- and many-photon regimes, fully
incorporating both nonlinearity and retardation effects \cite%
{PhysRevA.92.053834,PhysRevA.110.033707}. For nonzero $\phi $, input photons
from one direction can experience complete transparency---propagating
coherently without interacting with the system \cite{PhysRevX.13.021039}.
Conversely, when driven from the opposite direction, we find that the same
system generates output with strongly nonclassical photon statistics.

In the weak-driving regime, the emergence of non-Poissonian photon
statistics is governed by the system's intrinsic nonlinearity, while their
deterministic generation is protected by $\mathcal{PT}$ symmetry.
Surprisingly, even when the nonlinear scale is smaller than the decay
rate---deep in the dissipative regime---quantum interference within the
nonlocal coupling region leads to highly nontrivial correlations \cite%
{PhysRevLett.117.203602,PhysRevResearch.7.L022024}. These correlations are
direction-dependent \cite{PhysRevA.110.023722} and sensitive to the spatial
separation of the coupling points, reflecting the inherent non-Markovianity
of the system. Specifically, at $\phi =\pm \mathrm{\pi }/2$ \cite%
{PhysRevLett.133.063603}, interference between coupling points vanishes, and
the system's dynamics become exactly Markovian. This Markovian behavior is
not limited to the single-excitation case \cite{PhysRevLett.133.063603}, but
persists regardless of the excitation number. In this limit, a master
equation description becomes exact \cite{Gardiner2004,Walls2008}, allowing
for full characterization under arbitrary driving. Yet, the output fields
remain non-Markovian due to their interaction with the extended coupling
structure. This interplay between internal Markovianity and external memory
effects enables controlled realization of nonreciprocal dissipative phase
transitions (DPTs) \cite%
{houck2012chip,PhysRevA.86.012116,diehl2008quantum,PhysRevX.5.031028,PhysRevX.7.011016,RevModPhys.86.1391,PhysRevA.93.023846,PhysRevA.98.042118}%
. By tuning the chiral phase and leveraging nonlocal interference, our
framework establishes a versatile platform for engineering nonreciprocal,
non-Markovian multiphoton dynamics in waveguide QED.

\section{Results}

\begin{figure}[tbp]
\begin{center}
\includegraphics[width=0.9\linewidth]{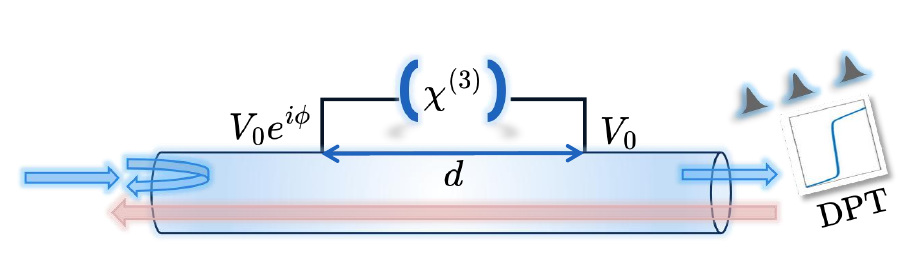}
\end{center}
\caption{\textbf{Schematic of a giant Kerr cavity coupled to a waveguide at
two coupling points separated by a distance }$d$\textbf{. }The coupling
phase $\protect\varphi $ can be tuned to a nonzero value, breaking parity
symmetry. When the propagation phase at the driving frequency matches $%
\protect\varphi $, the giant system interacts only with the right-moving
driving field (indicated by the right arrow), which is either reflected or
transmitted, while it remains transparent to the left-moving input
(indicated by the red arrow). In the former case, the transmitted light
reveals distinctive characteristics, such as nontrivial statistical
behaviors and the occurrence of dissipative phase transitions (DPTs).}
\label{fig1}
\end{figure}

As a paradigmatic example, we use a giant Kerr cavity \cite%
{PhysRevLett.79.1467,PhysRevLett.103.150503,RevModPhys.93.025005}, which
reduces to a giant two-level atom as its nonlinearity $U$ approaches
infinity, to demonstrate the multiphoton chiral dynamics. But it can also be
extended to other giant systems. The setup, illustrated in Fig. \ref{fig1},
involves a giant Kerr cavity nonlocally coupled to a 1D bath (the waveguide)
at two points, $0$ and $d$. Without loss of generality, we introduce a phase
shift $\varphi $ at coupling point $0$, as only the phase difference between
the two coupling points is relevant. The full Hamiltonian $H=H_{0}+H_{1}$,
where%
\begin{equation}
H_{0}=\frac{U}{2}b^{\dag 2}b^{2}+H_{\lambda \mathrm{D}}+\sum_{\lambda =%
\mathrm{L,R}}\int_{-\infty }^{+\infty }\varepsilon _{\lambda k}a_{\lambda
k}^{\dag }a_{\lambda k}\mathrm{d}k,
\end{equation}%
is the free Hamiltonian of the cavity and the waveguide respectively, plus
the driving term $H_{\lambda \mathrm{D}}=\left( \Omega _{\lambda }\left(
k_{i}\right) b^{\dagger }e^{-i\varepsilon _{\lambda k_{i}}t}+H.c.\right) $,
where $b$ ($a_{\lambda k}$) is the annihilation operator for the cavity mode
($\lambda $-moving bath mode with momentum $k$), and their interaction is
depicted by%
\begin{equation}
H_{1}=\sum_{\lambda =\mathrm{L,R}}\int_{-\infty }^{+\infty }V_{\lambda
k}a_{\lambda k}b^{\dag }\mathrm{d}k+\mathrm{H.c..}  \label{1}
\end{equation}%
Here, $\Omega _{\lambda }\left( k_{i}\right) =\Omega _{0}\eta _{\lambda
k_{i}}\left( \phi \right) $ is the Rabi frequency of the driving field
propagating in direction $\lambda $ with momentum $k_{i}$, where $\Omega
_{0} $ is a constant strength, and $\eta _{\lambda k_{i}}\left( \phi \right)
=e^{i\phi }+e^{i\theta _{\lambda k_{i}}}$, with the propagation phase $%
\theta _{\lambda k_{i}}=\left( \sigma _{\lambda }k_{0}+k_{i}\right) d$, $%
\sigma _{R}=1$, and $\sigma _{L}=-1$. The Hamiltonian is written in a
rotating frame with respect to the central frequency $k_{0}$ of the bath,
which is assumed to coincide with the cavity frequency for simplicity. The $%
k $-dependent coupling strength $V_{\lambda k}=V_{0}\eta _{\lambda k}\left(
\phi \right) $, where $V_{0}$ is a constant strength. The group velocity is
taken to be unity. The spectrum of the bath is assumed to be linear, i.e., $%
\varepsilon _{\lambda k}=\sigma _{\lambda }k$, but it can be extended to
other spectra, such as cosine spectrum \cite%
{PhysRevA.83.013825,PhysRevX.6.021027} and Ohmic spectrum \cite%
{PhysRevLett.120.153602}.

By tuning the phase $\phi $ to match the propagation phase at a driving
frequency $k_{i}$, corresponding to $\Omega _{\mathrm{L}}\left(
-k_{i}\right) $ and $\Omega _{\mathrm{R}}\left( k_{i}\right) $, the giant
system can be selectively decoupled from either the left- or right-moving
driving field, effectively realizing the chiral light-emitter interaction.
Additionally, reversing the direction of the input photons is equivalent to
changing $\phi $ to $-\phi $. Without loss of generality, in the following,
we focus on the case with
\begin{equation}
\phi =\left( 2n+1\right) \mathrm{\pi }-\left( k_{0}+k_{i}\right) d,
\label{4}
\end{equation}%
which results in $\Omega _{L}\left( -k_{i}\right) =0$, where $n$ is an
integer, and study the chiral dynamics for right-moving inputs with
frequency $k_{i}$.

\subsection{Weak-driving limit}

In the weak-driving limit, the dynamics are governed by the lowest-order
behavior of $\Omega _{0}$ and can be linked to few-photon scattering
processes \cite{PhysRevA.92.053834,PhysRevLett.117.203602}. For instance,
reflection and transmission probabilities can be acquired by calculating the
single-photon scattering processes, while second-order coherence can be
inferred from the two-photon wavefunction. In the case of single-photon
scattering, the reflection is reciprocal due to the $\mathcal{PT}$ symmetry
\cite{Sakurai_Napolitano_2020,PhysRevA.110.023722} of the system. Namely,
the reflection coefficients satisfy $\left\langle 0\right\vert a_{\mathrm{R}%
k}Sa_{\mathrm{L}-k_{i}}^{\dagger }\left\vert 0\right\rangle =\left\langle
0\right\vert a_{\mathrm{L}-k}Sa_{\mathrm{R}k_{i}}^{\dagger }\left\vert
0\right\rangle $, where $|0\rangle $ is the vacuum state and the $S$ matrix
is defined as $S=\lim_{T\rightarrow +\infty }^{\Omega _{0}\rightarrow
0}e^{iH_{0}T}e^{-i2HT}e^{iH_{0}T}$ \cite{Sakurai_Napolitano_2020}.
Consequently, the transmission probabilities for single left-moving and
right-moving input photons are identical, since no other dissipation
channels are present. Following the approach of \cite{PhysRevA.92.053834},
where the bath field is integrated out using a path-integral method with the
coherent-state representation (see Supplementary Note 1), we obtain the
self-energy \cite{fetter2012quantum,PhysRevX.6.021027}

\begin{eqnarray}
\Sigma \left( k\right)  &\equiv &\sum_{\lambda =\mathrm{L,R}}\int \frac{%
|V_{\lambda k}|^{2}\mathrm{d}k}{\omega -\varepsilon _{\lambda k}+i0^{+}}
\notag \\
&=&-i2\gamma \left( 1+\cos \phi e^{i\theta _{\mathrm{R}k}}\right) ,
\end{eqnarray}%
where $\gamma =2\mathrm{\pi }V_{0}^{2}$ is the decay rate accounting for
cavity-bath coupling at a single point. The real part of $\Sigma \left(
k\right) $ is the energy correction to the giant system, while its imaginary
part gives to the decay rate.\ Consequently, the reflection and transmission
coefficients with input and output momenta $k_{i}$ and $k_{\mathrm{f}}$, can
be written as (see Supplementary Note 2) $\left\langle 0\right\vert a_{%
\mathrm{L}k_{\mathrm{f}}}Sa_{Rk_{i}}^{\dagger }\left\vert 0\right\rangle
=r\left( k_{i}\right) \delta \left( k_{i}+k_{\mathrm{f}}\right) $ and $%
\left\langle 0\right\vert a_{Rk_{\mathrm{f}}}Sa_{Rk_{i}}^{\dagger
}\left\vert 0\right\rangle =t\left( k_{i}\right) \delta \left( k_{i}-k_{%
\mathrm{f}}\right) $, where%
\begin{equation}
r\left( k\right) =-i\gamma \eta _{\mathrm{L}-k}^{\ast }\left( \phi \right)
G\left( k\right) \eta _{\mathrm{R}k}\left( \phi \right)
\end{equation}%
and%
\begin{equation}
t\left( k\right) =1-i\gamma \left\vert \eta _{\mathrm{R}k}\left( \phi
\right) \right\vert ^{2}G\left( k\right) ,
\end{equation}%
with the Green function \cite%
{fetter2012quantum,PhysRevA.92.053834,PhysRevX.6.021027} $G\left( k\right) =%
\left[ k-\Sigma \left( k\right) \right] ^{-1}$.

Under the condition given by Eq. (\ref{4}), the $\mathcal{PT}$ symmetry of
the system ensures that the transmission probability for single right-moving
photons is unity, indicating deterministic photon generation in the
transmitted direction for both input orientations. However, the statistical
properties of the right-moving transmitted photons can markedly differ from
the trivial one for those ones moving to the left. Specifically, for two
input photons with identical momentum $k_{i}$, the $S$-matrix element $%
S_{p_{2}p_{1},k_{i}k_{i}}\equiv \left\langle 0\right\vert a_{\mathrm{R}%
p_{2}}a_{\mathrm{R}p_{1}}Sa_{\mathrm{R}k_{i}}^{\dagger }a_{\mathrm{R}%
k_{i}}^{\dagger }\left\vert 0\right\rangle $ governing the transmission of
two photons with momenta $p_{1}$ and $p_{2}$ is (see Supplementary Note 3)
\begin{eqnarray}
S_{p_{2}p_{1},k_{i}k_{i}} &=&2t\left( k_{i}\right) ^{2}\delta \left(
k_{i}-p_{1}\right) \delta \left( k_{i}-p_{2}\right)   \notag \\
&&-i\Gamma _{p_{2}p_{1},k_{i}k_{i}}\left( \phi \right) \delta \left(
2k_{i}-p_{1}-p_{2}\right)   \label{5}
\end{eqnarray}%
where
\begin{equation}
\Gamma _{p_{2}p_{1},k_{i}k_{i}}=\frac{\gamma ^{2}}{\pi }\eta _{\mathrm{R}%
p_{2}}^{\ast }\eta _{\mathrm{R}p_{1}}^{\ast }G\left( p_{1}\right) G\left(
p_{2}\right) T_{S}G^{2}\left( k_{i}\right) \eta _{\mathrm{R}k_{i}}^{2}
\end{equation}%
and%
\begin{equation}
T_{S}=\frac{U}{1-\frac{U}{k_{i}+i2\gamma }\sum_{mn}F_{mn}\left( k_{i}\right)
\Theta \left( m-n\right) }
\end{equation}%
depends on the input momentum $k_{i}$. The function
\begin{eqnarray}
F_{mn}\left( k\right)  &=&\frac{e^{i\left( m-n\right) \left( k+i2\gamma
\right) d}}{m!}\times   \notag \\
&&\left( \frac{-i\gamma \cos \phi e^{i\theta _{\mathrm{R}k}}}{k+i2\gamma }%
\right) ^{m+n}\sum_{l=0}^{n}C_{mnl}\left( k\right) ,
\end{eqnarray}%
with $C_{mnl}\left( k\right) =\frac{\left( m+n-l\right) !}{\left( n-l\right)
!l!}\left[ -i2\left( m-n\right) \left( k+i2\gamma \right) d\right] ^{l}$. In
Eq. (\ref{5}), the first term is resulted from the independent scattering of
the two photons, while the second term represents the background
fluorescence \cite{PhysRevLett.98.153003,PhysRevB.79.205111}.

The second-order correlation function%
\begin{equation}
g_{\lambda }^{\left( 2\right) }\left( \tau \right) \equiv \frac{\left\langle
a_{out}^{\lambda \dagger }\left( 0\right) a_{out}^{\lambda \dagger }\left(
\tau \right) a_{out}^{\lambda }\left( \tau \right) a_{out}^{\lambda }\left(
0\right) \right\rangle }{\left\langle a_{out}^{\lambda \dagger }\left(
0\right) a_{out}^{\lambda }\left( 0\right) \right\rangle ^{2}},
\end{equation}%
where the output fields $a_{out}^{\lambda }\left( t\right) =\frac{1}{\sqrt{%
2\pi }}\int a_{\lambda k}\left( T\right) e^{-i\sigma _{\lambda }k\left(
t-T\right) }\mathrm{d}k$, is connected to the two-photon wavefunction $%
w_{\lambda }\left( X_{1},X_{2}\right) $ with two photons at positions $X_{1}$
and $X_{2}$ as $g_{\lambda }^{\left( 2\right) }\left( \tau \right)
=\left\vert \frac{w_{\lambda }\left( \tau ,0\right) }{w_{\lambda }\left(
\tau \rightarrow \infty ,0\right) }\right\vert ^{2}$, while\ the
wavefunction $w_{t}\left( X_{1},X_{2}\right) $ of two transmitted photons
can be acquired by performing the Fourier transform of $%
S_{p_{2}p_{1},k_{2}k_{1}}$ as%
\begin{equation}
w_{t}\left( X_{1},X_{2}\right) =\frac{1}{2\pi }\int
dp_{1}dp_{2}e^{ip_{1}X_{1}}e^{ip_{2}X_{2}}S_{p_{2}p_{1},k_{2}k_{1}}.
\end{equation}

\begin{figure}[tbp]
\begin{center}
\includegraphics[width=0.9\linewidth]{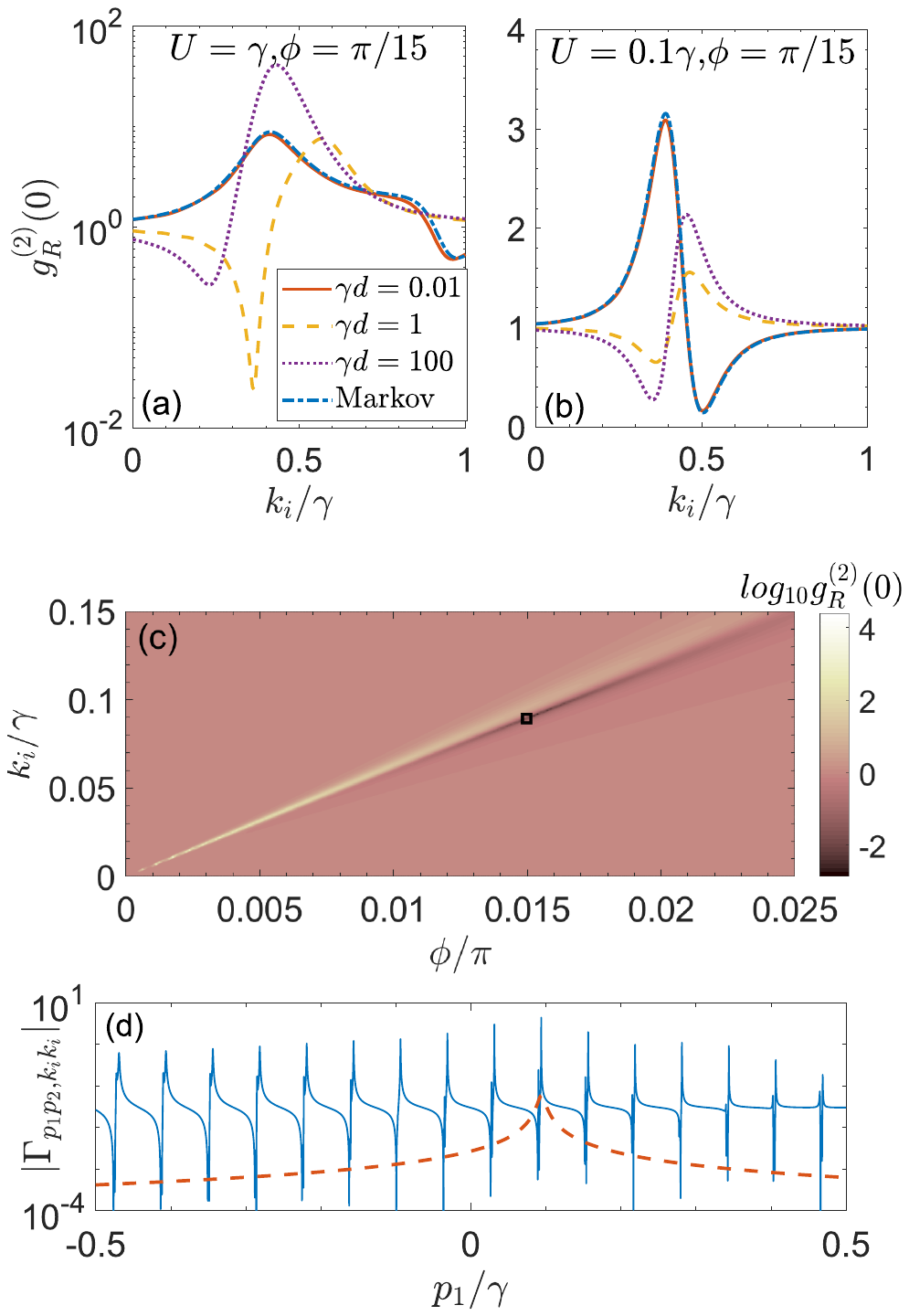}
\end{center}
\caption{\textbf{Properties of the output photons in the weak-driving limit.}%
\textrm{\ }(a) Second-order correlation function $g_{R}^{(2)}(0)$ of the
transmitted light with $\protect\phi =\protect\pi /15$ and $U=\protect\gamma
$. (b) Same as (a), but for weak nonlinearity $U=0.1\protect\gamma $. (c) $%
g_{R}^{(2)}(0)$ as a function of $k_{i}$ and $\protect\phi $ in the strong
dissipation limit with $U=0.01\protect\gamma $ and $\protect\gamma d=100.$
(d) Background fluorescence $\left\vert \Gamma
_{p_{2}p_{1},k_{i}k_{i}}\right\vert $ with parameters from the rectangle in
(c) shown in blue solid line. Here, in $\Gamma _{p_{2}p_{1},k_{i}k_{i}}$, $%
p_{2}=2k_{i}-p_{1}$. For comparison, $\left\vert \Gamma
_{p_{2}p_{1},k_{i}k_{i}}\right\vert $ in the Markovian regime with $\protect%
\gamma d=0.01$ is shown as the red dashed curve. }
\label{fig2}
\end{figure}

By fixing $\theta _{\mathrm{R}k_{i}}=\mathrm{\pi }-\phi $, we investigate
the statistics of transmitted photons for right-moving inputs, varying both $%
k_{i}$ and $\phi $. The results are depicted in Fig. \ref{fig2}. When $U$ is
comparable to the decay rate $\gamma $, the output photons can exhibit
sub-Poissonian ($g_{\mathrm{R}}^{(2)}(0)<1$) or super-Poissonian ($g_{%
\mathrm{R}}^{(2)}(0)>1$) statistics for input momenta $k_{i}$ within a width
approximately $\gamma $. In contrast to the single-point coupling scenario
\cite{PhysRevA.84.063803}, nontrivial statistical behaviors of transmitted
photons persist even in the strong dissipation regime where $\gamma \gg U$,
as shown in Fig. \ref{fig2}(b).

In the Makovian regime with $\gamma d\ll 1$, the second-order correlation
function%
\begin{equation}
g_{\mathrm{R}}^{\left( 2\right) }\left( 0\right) =\left\vert \tilde{t}\left(
k_{i}\right) ^{2}+4\gamma _{0}^{2}\tilde{G}\left( k_{i}\right) ^{2}-\frac{%
\gamma _{0}^{2}\tilde{G}\left( k_{i}\right) }{k_{i}-h}\right\vert ^{2},
\label{6}
\end{equation}%
where the transmission amplitude $\tilde{t}\left( k_{i}\right) =1-2i\gamma
_{0}\tilde{G}\left( k_{i}\right) $, with$\ \tilde{G}\left( k\right) =\left[
k-\Sigma \left( 0\right) \right] ^{-1}$, $\gamma _{0}\equiv -$Im$\left[
\Sigma (0)\right] $ is the effective decay rate of the giant cavity, and $%
h\equiv \Sigma \left( 0\right) -U/2$ is half of the frequency of two
excitations. In Eq. (\ref{6}), the first term inside the absolute value
arises from independent photon scattering, while the last two terms
originate from the correlated scattering processes. The photon statistics of
the transmitted light are determined by the interference between these
different contributions. The last term represents the simultaneous
absorption of two photons, which becomes negligible when $U$ is sufficiently
large.\ When $\phi =\pi /15$, the self-energy is $\Sigma (0)=\left(
0.4-i0.09\right) \gamma $, and the input photon frequency $k_{i}=0.5\gamma $
is close to the remormalized single-photon energy $k_{0}$. In the large $U$
limit, the two-photon absorption process is strongly suppressed, so the main
contribution to $g_{\mathrm{R}}^{\left( 2\right) }\left( 0\right) $ comes
from the first two terms that can be combined as $\frac{k_{i}-k_{0}-i3\gamma
_{0}}{k_{i}-k_{0}+i\gamma _{0}}$, with $k_{0}=$Re$\left[ \Sigma (0)\right] $%
, resulting in super-Poissonian statistics, as shown in Fig. \ref{fig2}(a).
This also explains why $g_{R}^{\left( 2\right) }\left( 0\right) $ reaches
its maximum at resonant input $k_{i}=0.4\gamma $ in Fig. \ref{fig2}(a), as
well as in Fig. 2(b) at the same frequency. In contrast, when $U=0.1\gamma $%
, the input frequency $k_{i}=0.5\gamma $ is close to two photon resonance
condition, where $h$ is small. This enhances the contribution from the
two-photon absorption process. Under the parameters used in Fig. \ref{fig2}%
(b), this contribution largely cancels the first term, leading to
sub-Poissonian photon statistics. As the distance $d$ increases, the system
enters the non-Markovian regime, where the photon statistics can differ
significantly from the Markovian case.

Further tuning of $k_{i}$ and $\varphi $ reveals that the second-order
correlation function $g_{\mathrm{R}}^{(2)}(0)$ varies in a wide range from $%
10^{-2}$ to $10^{4}$ (Fig. \ref{fig2}(c)). This nontrivial statistical
behavior arises from photon interference between the two coupling points,
resulting in an effective decay rate $\gamma _{k}\equiv -$Im$\left[ \Sigma
(k)\right] =2\gamma \left( 1+\cos \phi \cos \theta _{\mathrm{R}k}\right) $
for the $k$-th mode, which can be significantly smaller than $U$. These
modes can be efficiently excited in the two-photon scattering process by
matching its momentum $k$ with the real part of the
self-energy, $\text{Re}\left[ \Sigma (k)\right] =2\gamma \cos \phi \sin
\theta _{\mathrm{R}k}$ or $2k_{i}-$Re$\left[ \Sigma (2k_{i}-k)\right] $.
Fig. \ref{fig2}(d) illustrates this phenomenon with the background
fluorescence \cite{PhysRevLett.98.153003,PhysRevB.79.205111} $\Gamma
_{p_{2}p_{1},k_{i}k_{i}}$ at the point indicated by the rectangle in Fig. %
\ref{fig2}(c), where $k_{i}=0.09\gamma $, $\phi =0.015\pi $, and $g_{\mathrm{%
R}}^{(2)}(0)=0.017$, corresponding to $\Sigma (k_{i})=(0.094-i0.0044)\gamma $%
. The blue solid curve in Fig. \ref{fig2}(d) shows highly populated modes
centered around $0.094+2n\pi /d$ with a narrow width of $\gamma _{k}/\left(
\gamma L\right) $, underscoring the nontrivial statistical behaviors of
transmitted photons. In contrast, when $d$ decreases, the number of modes
satisfying $k=$Re$\left[ \Sigma (k)\right] $ decrease, leading to less
resonant peaks in the background fluorescence, together with larger width $%
\gamma _{k}$. Specifically, as shown in the red dashed curve, in the
Makovian regime with $\gamma d=0.01$, the equation $k=$Re$\left[ \Sigma (k)%
\right] $ has only one resonant peak, resulting in different statistics of
output photons.

We note that throughout the manuscript, the frequency of the cavity is set
equal to the central frequency $k_{0}$ of the waveguide. In a gernal case, a
detuning $\Delta _{c}$ between the cavity frequency and $k_{0}$ can be
introduced. With this detuning, all occurrences of $k$ in the
expressions---except in the phase factor $\theta _{\lambda k}$---should be
replaced by $k-\Delta _{\mathrm{c}}$. By redefining $\tilde{k}=k-\Delta _{%
\mathrm{c}}$, the phase $\theta _{\mathrm{R}k}=\left( k_{0}+\tilde{k}%
_{i}+\Delta _{c}\right) d$, and the chirality condition in Eq. (\ref{4}) is
modified to%
\begin{equation}
\phi =\left( 2n+1\right) \mathrm{\pi }-\left( k_{0}+\tilde{k}_{i}+\Delta
_{c}\right) d,
\end{equation}%
Since $\theta _{\mathrm{R}k}=\left( 2n+1\right) \pi -\phi $, this
redefinition allows all other expressions to retain their original form with
$k$ replaced directly by $\tilde{k}$. Thus, when specifying $\tilde{k}_{i}$
as the input momentum, the results remain unchanged. For given values of $%
\phi $, $k_{0}$, and $d$, the chirality condition can be satisfied for
different $\tilde{k}_{i}$ by tuning $\Delta _{\mathrm{c}}$.

\subsection{Markovianity at $\protect\phi =\pm \mathrm{\protect\pi }/2$}

When $\gamma d\ll 1$, the two-photon dynamics simplify to be Markovian
(dashed-dotted line in Figs. \ref{fig2}(a) and (b)), and the system's
evolution can be approximatly described by a master equation and the
input-output formalism \cite{Gardiner2004,Walls2008}. Generally, this
Markovian approximation breaks down for large distances $d$, as evidenced by
the deviation in the second-order correlator $g_{\mathrm{R}}^{(2)}(0)$
between $\gamma d=0.01$ and $\gamma d=1,100$ shown in Figs. \ref{fig2}(a)
and (b)\ (for fixed $\theta _{\mathrm{R}k_{i}}$, Markovian dynamics are
independent of $d$). However, when $\phi =\pm \mathrm{\pi }/2$, the
self-energy $\Sigma (k)$ becomes independent of $k$, and the evolution of
the giant cavity's density matrix $\rho $ can be exactly described by the
master equation \cite{PhysRevA.92.053834} regardless of the distance $d$ and
photon number:
\begin{equation}
\partial _{t}\rho =-i\left[ H_{S},\rho \right] +2\gamma \left( 2b\rho
b^{\dagger }-\left\{ b^{\dagger }b,\rho \right\} \right) ,  \label{2}
\end{equation}%
where $\rho $ is the density matrix of the cavity and $H_{S}=\frac{U}{2}%
b^{\dag 2}b^{2}+H_{\lambda D}$. In this scenario, the decay of the cavity
due to the interference between right- and left-moving photons propagating
between the two coupling points is completely canceled out.

This cancellation can also be understood by defining two modes
\begin{equation}
\left(
\begin{array}{c}
A_{+k} \\
A_{-k}%
\end{array}%
\right) =U_{k}\left(
\begin{array}{c}
a_{L-k} \\
a_{Rk}%
\end{array}%
\right) ,
\end{equation}%
where the unitary transformation $U_{k}=\frac{1}{\tilde{V}_{k}}\left(
\begin{array}{cc}
V_{\mathrm{L}-k} & V_{\mathrm{R}k} \\
V_{\mathrm{R}k}^{\ast } & -V_{\mathrm{L}-k}^{\ast }%
\end{array}%
\right) $, with $\tilde{V}_{k}=2V_{0}\sqrt{1+\cos \phi \cos \left(
k_{0}+k\right) L}$. These two modes are degenerated and decoupled, and the
cavity field only couples to $A_{+k}$ mode with the interaction in Eq. (\ref%
{1}) rewritten as%
\begin{equation}
H_{1}=\int_{-\infty }^{+\infty }\tilde{V}_{k}\left( b^{\dag }A_{+k}+\mathrm{%
H.c.}\right) \mathrm{d}k\mathrm{.}
\end{equation}%
When $\phi =\pm \mathrm{\pi }/2$, $\tilde{V}_{k}$ is independent of $k$,
leading to a constant decay rate $2\gamma $ contributed from the incoherent
superposition of cavity photons decaying via the two coupling points. Note
that this $k$-independence in $\tilde{V}_{k}$ with $\phi =\pm \pi /2$ is
general and does not depend on the specific spectrum $\varepsilon _{\lambda
k}$. However, the Markovian description is rigorously valid only for a
linear spectrum. Furthermore, the properties of photons propagating in the
waveguide can be acquired with the time-nonlocal input-output relation%
\begin{equation}
a_{\mathrm{out}}^{\lambda }\left( t\right) =a_{\mathrm{in}}^{\lambda }\left(
t\right) -i\sqrt{\gamma }\left[ e^{-i\phi }b\left( t\right) +e^{-i\theta
_{\lambda 0}}b\left( t+\sigma _{\lambda }d\right) \right] ,  \label{3}
\end{equation}%
where the input field $a_{\mathrm{in}}^{\lambda }\left( t\right) =\frac{1}{%
\sqrt{2\pi }}\int a_{\lambda k}\left( 0\right) e^{-i\sigma _{\lambda }kt}%
\mathrm{d}k$. Note that unlike the giant system itself, the output field $a_{%
\mathrm{out}}^{\lambda }\left( t\right) $ is determined by the cavity field
at two distinct times $t$ and $t\pm d$ even when $\phi =\pm \mathrm{\pi }/2$%
. This indicates that the non-Markovian characteristic is preserved in the
properties of the output photons.

\subsection{Non-Makovian dissipative phase transition}

\begin{figure}[tbp]
\begin{center}
\includegraphics[width=1\linewidth]{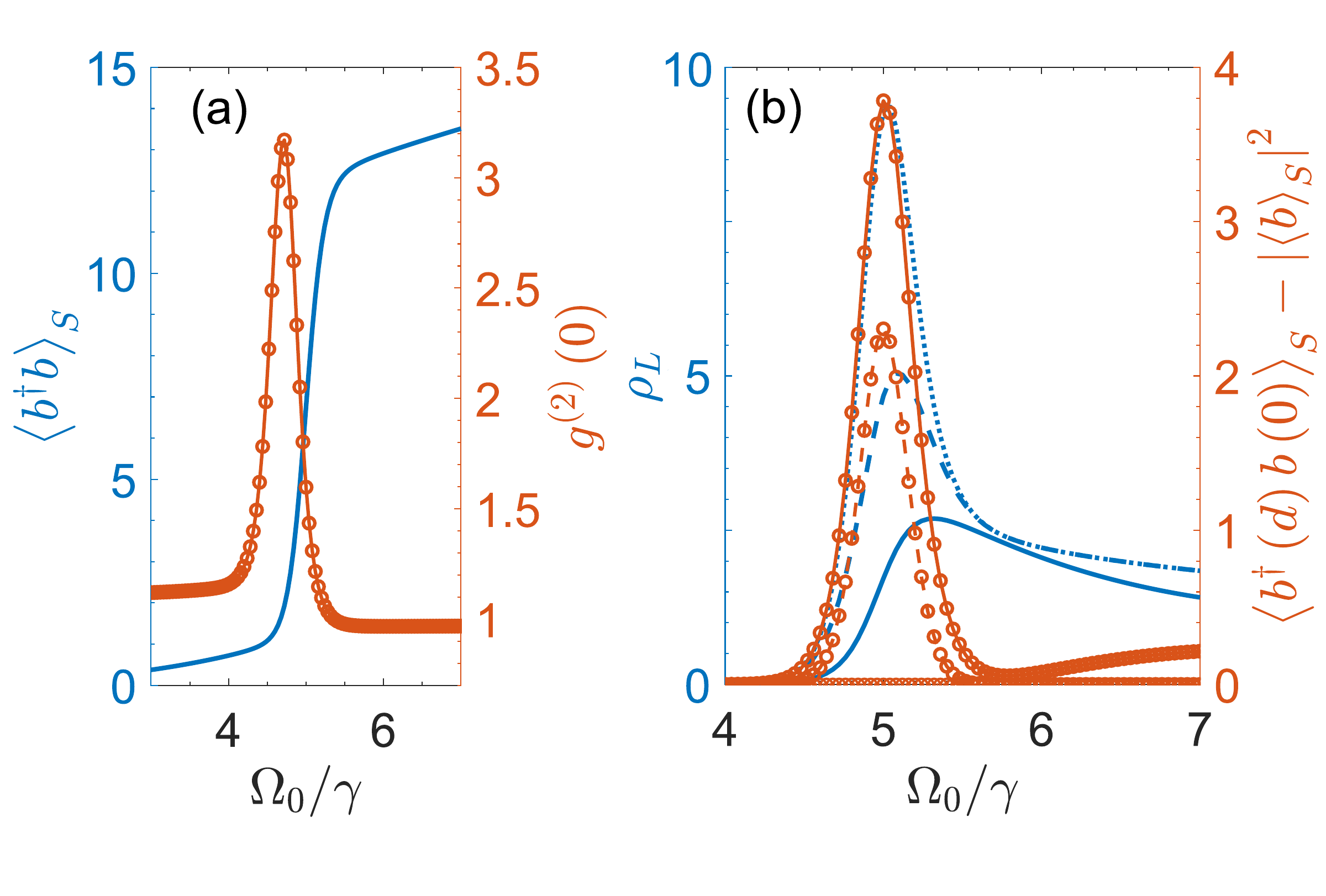}
\end{center}
\caption{\textbf{Dissipative phase transition occurs only when the giant
system is driven by right-moving input.} (a) Photon number (blue solid line,
left axis) in the cavity and the second-order correlation function (red line
with circles, right axis), as functions of the driving strength (b) Photon
number density (blue lines, left axis) of the reflected field and the
two-point correlation (red lines with circles, right axis) for different
distance $\protect\gamma d=0.63$ (solid lines), $\protect\gamma d=63$
(dashed lines), and $\protect\gamma d=6300$ (dotted lines). Other parameters
are $U=\protect\gamma $ and $k_{i}=10\protect\gamma $.}
\label{fig3}
\end{figure}

The master equation (\ref{2}), along with the input-output formalism (\ref{3}%
), enables a comprehensive exploration of the giant system's dynamics and
the characteristics of the output photons under arbitrarily strong driving
fields. When $\phi =(k_{0}+k_{i})d=\mathrm{\pi }/2$, the giant system
interacts exclusively with the right-moving driving field, leading to the
emergence of a DPT in the strong-driving regime \cite%
{drummond1980quantum,drummond1980generalised,PhysRevA.94.033841}. As
depicted in Fig. \ref{fig3}(a), increasing the pump strength $\Omega _{0}$
results in a sudden increase in the cavity photon number $\langle b^{\dag
}b\rangle _{\mathrm{S}}$ around $\Omega _{0}\approx 5\gamma $, corresponding
to a peak in the second-order correlator $g^{(2)}(0)\equiv \frac{\langle
b^{\dag 2}b^{2}\rangle _{\mathrm{S}}}{\langle b^{\dag }b\rangle _{\mathrm{S}%
}^{2}}$, where $\langle ...\rangle _{\mathrm{S}}$ denotes the steady-state
average. Here, we consider the Kerr nonlinearity $U=\gamma $ and incident
momentum $k_{i}=10\gamma $. Using methods such as the complex
P-representation formalism \cite%
{Walls2008,drummond1980quantum,drummond1980generalised,PhysRevA.94.033841}
or the quantum absorber approach \cite{Stannigel_2012,PhysRevX.10.021022},
one can analytically derive the steady-state solution of the master equation
(\ref{2}), revealing a first-order DPT in the thermodynamic limit where the
Liouvillian gap (the real part of the second-largest eigenvalue of the
Liouvillian) vanishes at the critical point \cite{PhysRevA.98.042118}.

This DPT-induced sudden change manifests not only in the cavity field but
also in the reflected field, highlighting the non-Markovian effects induced
by the giant system. At $T\rightarrow \infty $, the density $\rho _{L}\equiv
\frac{1}{2\mathrm{\pi }}\int \langle a_{Lk_{2}}^{\dag
}(T)a_{Lk_{1}}(T)\rangle e^{i(k_{1}-k_{2})X}\mathrm{d}k_{1}\mathrm{d}k_{2}$
of left-moving output photons at position $X<0$ can be computed using the
input-output relation (\ref{3}) as
\begin{eqnarray}
\rho _{L} &=&\left\langle a_{\mathrm{out}}^{\mathrm{L}\dag }\left(
T+x\right) a_{\mathrm{out}}^{\mathrm{L}}\left( T+x\right) \right\rangle
\notag \\
&=&2\gamma \left( \left\langle b^{\dag }b\right\rangle _{\mathrm{S}}-\text{Im%
}\left[ e^{ik_{0}d}\left\langle b^{\dag }\left( d\right) b\left( 0\right)
\right\rangle _{\mathrm{S}}\right] \right) .  \label{7}
\end{eqnarray}%
The DPT observed in the reflected field can differ significantly from that
of the cavity field. For instance, under the conditions $\phi =\pm \pi /2$
and Eq. (\ref{4}), the phase transition in the cavity field is independent
of $d$. In contrast, the critical behavior in the reflected field is highly
sensitive to $d$ due to the nonlocal correlation $\left\langle b^{\dag
}(d)b(0)\right\rangle _{\mathrm{S}}$. When $d$ approaches zero, no reflected
photons are observed, and thus no phase transition occurs in the reflected
field. As $d$ increases, the DPT in the reflected field emerges, revealing
the non-Markovian nature of the critical behavior in the output field.

To quantitatively assess the impact of non-Markovian effect on the DPT, we
set $k_{0}d=\pi /2$ and $d=2n\pi /k_{i}$ to satisfy the chirality condition
in Eq. (\ref{4}), and examine the DPT via $\rho _{L}$ for various values of $%
d$. As $d$ increases, the DPT in the reflected field becomes more
pronounced, as illustrated in Fig. \ref{fig3}(b). Additionally, a peak
emerges near the critical point where the Liouvillian gap reaches a minimum
(vanishing only in the limit of infinite photon number), resulting in
maximal correlations. $\left\langle b^{\dag }(d)b(0)\right\rangle _{\mathrm{S%
}}-\left\vert \left\langle b\right\rangle _{\mathrm{S}}\right\vert ^{2}$. As
$d$ further increases, the peak in $\rho _{L}$ approaches the critical
point, and the magnitude of the correlator $\left\langle b^{\dag
}(d)b(0)\right\rangle _{\mathrm{S}}-\left\vert \left\langle b\right\rangle _{%
\mathrm{S}}\right\vert ^{2}$ diminishes because only the eigenstate with the
smallest gap contributes significantly. Specifically, when $d$ exceeds the
Liouvillian gap significantly, such as $d=2000\pi /k_{i}$, represented by
the dotted lines in Fig. \ref{fig3}(b), no correlations are present and $%
\left\langle b^{\dag }(d)b(0)\right\rangle _{\mathrm{S}}$ approaches $%
\left\vert \left\langle b\right\rangle _{\mathrm{S}}\right\vert ^{2}$. In
this scenario, the photon density $\rho _{L}$ reaches its maximum since $%
\left\langle b^{\dag }(d)b(0)\right\rangle _{\mathrm{S}}$ deviates most
significantly from $\left\langle b^{\dag }b\right\rangle _{\mathrm{S}}$.

It is also worth noting that the reflected field behaves differently from
the cavity field beyond the critical point. While the cavity photon number
continues to grow with the driving amplitude $\Omega _{0}$, the reflected
photon density $\rho _{L}$decreases after the critical point. This reduction
arises from the suppression of the first-order correlator $\left\langle
b^{\dag }(d)b(0)\right\rangle _{\mathrm{S}}-\left\vert \left\langle
b\right\rangle _{\mathrm{S}}\right\vert ^{2}$ after the transition. The
diminishing correlations are further evidenced by the second-order
correlation $g^{\left( 2\right) }\left( 0\right) $ approaching unity, as
shown in Fig. \ref{fig3}(a). After the critical point, $\left\langle b^{\dag
}(d)b(0)\right\rangle _{\mathrm{S}}\approx \left\vert \left\langle
b\right\rangle _{S}\right\vert ^{2}$ an $\rho _{L}\approx 2\gamma \left(
\left\langle b^{\dag }b\right\rangle _{\mathrm{S}}-\left\vert \left\langle
b\right\rangle _{\mathrm{S}}\right\vert ^{2}\right) $. Since both $%
\left\langle b^{\dag }b\right\rangle _{\mathrm{S}}$ and $\left\vert
\left\langle b\right\rangle _{\mathrm{S}}\right\vert ^{2}$ increase with the
driving amplitude $\Omega _{0}$, and the cavity field approaches a coherent
state in the large $\Omega _{0}$ limit, the reflected photon density $\rho
_{L}$ consequently decreases and eventually vanishes. These findings
demonstrate that this system offers a valuable platform for exploring
non-Markovian effects in DPTs.

\subsection{Connection to experimental realizations}

In this section, we present possible experimental implementations and
discuss the impact of small deviations from the condition in Eq. (\ref{4})
on the direction-dependent photon correlations and the nonreciprocal
features of the dissipative phase transition.

\subsubsection{Implementation}

Our theoretical proposal can be implemented in superconducting circuit
platforms. As demonstrated in a recent experiment \cite{PhysRevX.13.021039},
a giant system is coupled to a waveguide with a controllable phase
difference between two spatially separated coupling points. As shown in Fig. \ref%
{fig4}(a), two sets of auxiliary resonators $C$ and $R$ are used. $C$ are
tunable couplers, whose frequencies are modulated periodically via flux bias
lines. $R$ are used to filter undesire emission into the waveguide. The
phase difference $\phi _{\mathrm{L}}-\phi _{\mathrm{R}}$ corresponds to the
phase $\phi $ in our model.\ The giant Kerr cavity can be realized as an LC
circuit incorporating a nonlinear element like a Josephson junction \cite%
{RevModPhys.93.025005}, realizing a finite Kerr nonlinearity $U$.

\begin{figure}[tbp]
\begin{center}
\includegraphics[width=1\linewidth]{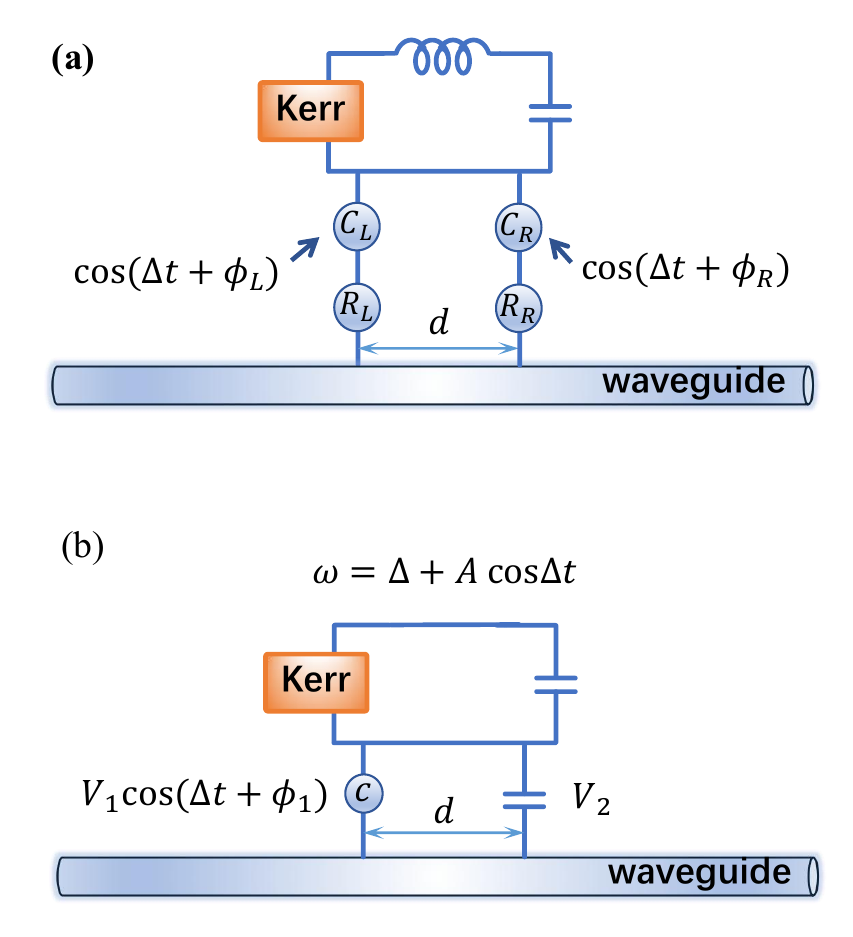}
\end{center}
\caption{\textbf{Schematics of the implementations of the theoretical model
using superconducting circuits.} In (a), four auxiliary resonators are
employed. The frequencies of two of them---the couplers ($C$)---are
periodically modulated with a phase difference $\protect\phi =\protect\phi %
_{L}-\protect\phi _{R}$. In (b), both the cavity frequency and its coupling
to the waveguide at one end are periodically modulated. The coupling
modulation is realized via a variable coupler composed of inductors and a
SQUID.}
\label{fig4}
\end{figure}

Alternatively, one can realize the phase difference by periodically
modulating both the frequency of the cavity and its coupling strength to the
waveguide. As shown in Fig. \ref{fig4}(b), at one end, the giant cavity couples
to the waveguide via a variable coupler made of inductors and a SQUID \cite%
{PhysRevLett.110.107001,PhysRevLett.113.220502,PhysRevB.87.134504,Wang_2022}%
. By modulating the flux, the coupling is modulated as $\cos \left( \Delta
t+\phi ^{\prime }\right) $, where $\Delta $ is the detuning of the cavity and%
$\ \phi ^{\prime }$ is the modulation phase. The Hamiltonian of the total
system $\tilde{H}\equiv H_{\mathrm{B}}+\tilde{H}_{\mathrm{S}}+\tilde{H}_{1}$%
, where%
\begin{equation}
\tilde{H}_{S}=\left( \Delta +A\cos \Delta t\right) b^{\dag }b+Ub^{\dag
2}b^{2}
\end{equation}%
is the giant system's Hamiltoian with modulation amplitude $A$ and%
\begin{eqnarray}
\tilde{H}_{1} &=&\int_{-\infty }^{+\infty }\mathrm{d}k[V_{1}\left( a_{%
\mathrm{L}k}+a_{\mathrm{R}k}\right) b^{\dag }\cos \left( \Delta t+\phi
^{\prime }\right) +  \notag \\
&&V_{2}\left( a_{\mathrm{L}k}e^{-i\left( k_{0}-k\right) L}+a_{\mathrm{R}%
k}e^{i\left( k_{0}+k\right) L}\right) b^{\dag }+\mathrm{H.c.}
\end{eqnarray}%
is its linear coupling to the waveguide, with coupling constants $V_{1}$ and
$V_{2}$. In the rotating frame with respect to $\tilde{H}_{0}\left( t\right)
=\left( \Delta +A\cos \Delta t\right) b^{\dag }b$, the Hamitltonian becomes%
\begin{equation}
H=U^{\dagger }\left( t,0\right) \tilde{H}\left( t\right) U\left( t,0\right) -%
\tilde{H}_{0}\left( t\right) \approx H_{0}+H_{1},
\end{equation}%
where the time evolution operator $U\left( t,0\right) =\mathcal{T}%
e^{-i\int_{0}^{t}\tilde{H}_{0}\left( t^{\prime }\right) dt^{\prime
}}=e^{-i\left( \Delta t+\frac{A}{\Delta }\sin \Delta t\right) b^{\dag }b}$.
Here, we have ignored the fast oscillation terms assuming that $\left\vert
V_{1,2}J_{n}\left( \frac{A}{\Delta }\right) \right\vert ^{2}\ll \left\vert
\Delta \right\vert $, and $V$ and $\phi $ are connected to the modulation
parameters as%
\begin{eqnarray}
V &=&\frac{V_{1}}{2}\left\vert J_{2}\left( \frac{A}{\Delta }\right) e^{i\phi
^{\prime }}+J_{0}\left( \frac{A}{\Delta }\right) e^{-i\phi ^{\prime
}}\right\vert  \notag \\
&=&V_{2}J_{1}\left( \frac{A}{\Delta }\right)
\end{eqnarray}%
and
\begin{equation}
\phi =\text{Arg}\left( J_{2}\left( \frac{A}{\Delta }\right) e^{i\phi
^{\prime }}+J_{0}\left( \frac{A}{\Delta }\right) e^{-i\phi ^{\prime
}}\right) .
\end{equation}

\subsubsection{Deviations from the chirality condition in (\protect\ref{4})}

\begin{figure}[tbp]
\begin{center}
\includegraphics[width=1\linewidth]{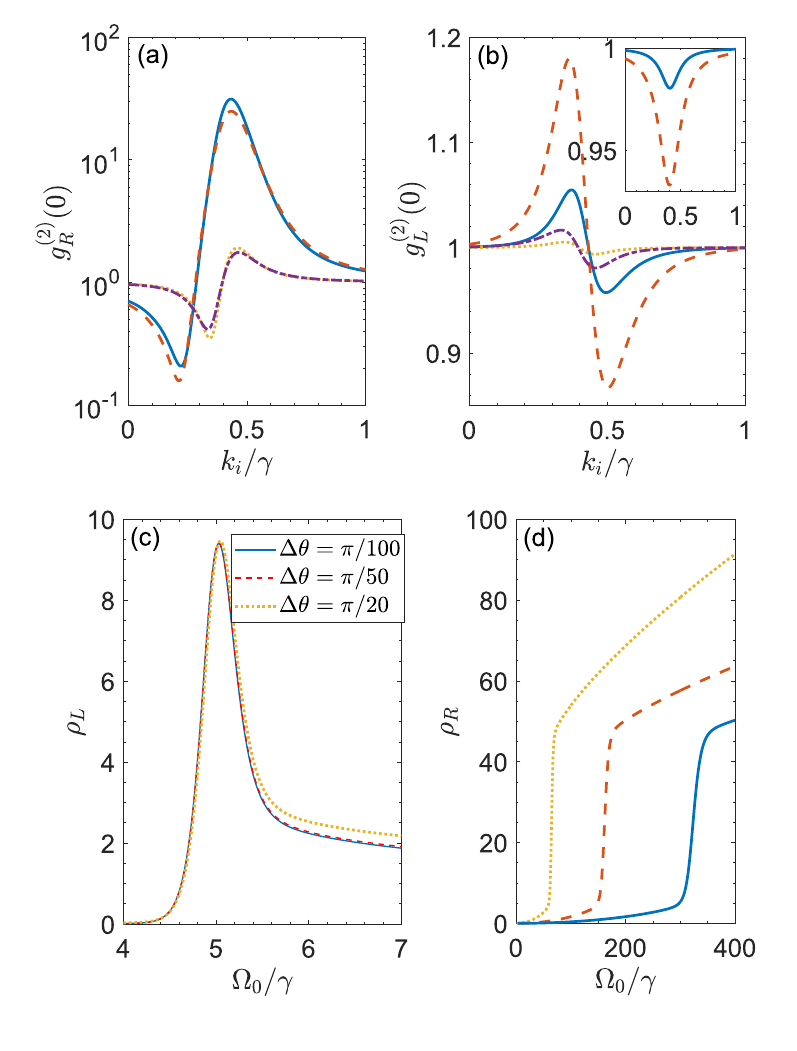}
\end{center}
\caption{\textbf{Deviations from the chirality condition.} Second-order
correlation function $g_{R,L}^{(2)}(0)$ of the transmitted photons in the
weak-driving limit, and the density of the reflected photons in the
strong-driving limit, when the chirality condition is slightly violated. The
driving field is right-moving in (a) and (c), and left-moving in (b) and
(d). In (a) and (b), the distance is $d=100/\protect\gamma $, with $U=%
\protect\gamma $ (blue solid and red dashed curves) and $0.1\protect\gamma $
(yellow dotted and purple dash-dotted curves). The phase deviations are $%
\Delta \protect\phi =\protect\pi /100$ (blue solid and yellow dotted curves)
and $\protect\pi /50$ (red dashed and purple dash-dotted curves). The inset
in (b) shows the transmission probability, with the legend consistent with
that in (a) and (b). In (c) and (d), the phase is fixed at $\protect\phi =%
\protect\pi /2$, the distance is $d=6300/\protect\gamma $, and different
deviations of the propagation phase are considered. Other parameters are the
same as in Fig.~\protect\ref{fig3}.}
\label{fig5}
\end{figure}

When the condition in Eq. (\ref{4}) is not fulfilled, the properties of the
transmitted photons under right-moving driving are altered, and the behavior
of transmitted photons under left-moving driving becomes nontrivial, since
the giant system can now be driven by the left-moving input. For a small
phase deviation $\Delta \phi $ from the value specified in Eq. (\ref{4}),
the second-order correlation functions $g_{R,L}^{\left( 2\right) }\left(
0\right) $ for the transmitted photons are shown in Figs. \ref{fig5}(a) and (b).
The phase is set to $\phi =\mathrm{\pi }/15+\Delta \phi $, with other
parameters the same as in Figs. \ref{fig2}(a) and (b).

Compare to the cases in Figs. \ref{fig2}(a) and (b) with right-moving
driving, the transmission probability is slight less than $1$, and the
second-order correlation function $g_{R}^{\left( 2\right) }\left( 0\right) $
shows only minor deviation for small $\Delta \phi =\pi /100$ and $\pi /50$.
Meanwhile, $g_{L}^{\left( 2\right) }\left( 0\right) $ under left-moving
driving exhibits nontrivial statistics, with $g_{L}^{\left( 2\right) }\left(
0\right) $ slightly deviating from 1. Notably, the deviations in $%
g_{R,L}^{\left( 2\right) }\left( 0\right) $ diminish for smaller $U$,
because only modes with smaller effective decay rate $\gamma _{k}$
contribute to nontrivial second-order correlation functions. For smaller $U$%
, fewer of these modes are effectively excited, resulting in a reduced
sensitivity to $\Delta \phi $. Therefore, for sufficiently small $\Delta
\phi $, $g_{L}^{\left( 2\right) }\left( 0\right) $ for left-moving driving
remains approximately 1, and nontrivial photon statistics are generated
predominantly for a single incident direction.

We now consider the impact of a small deviation from Eq. (\ref{4}) on the
DPT under the strong-diving limit. To ensure Markovian dynamics in the giant
system, $\phi $ is set to be $\mathrm{\pi }/2$, and a small deviation $%
\Delta \theta $ is introduced, making the propagation phase $\theta _{%
\mathrm{R}k_{i}}=\mathrm{\pi }/2+\Delta \theta $. With other parameters
identical to those in Fig. \ref{fig3}, the density of reflected photons
under right-moving driving shows negligible change, as illustrated in Fig. %
\ref{fig5}(c). Under left-moving driving, the reflected photon density $\rho
_{R}$ also exhibits a DPT, but at a critical pumping amplitude $\Omega _{0}$
that is more than 10 times larger, as shown in Fig. \ref{fig5}(d). This is
because the effective driving amplitude $\Omega _{L}\left( k_{i}\right) \ll
\Omega _{R}\left( k_{i}\right) $. As a result, if we focus on the critical
behavior within the driving amplitude range around the critical point for
right-moving driving, the DPT remains nonreciprocal.

\section{Conclusion}

In summary, we have explored the non-Markovian multiphoton dynamics of a
driven giant system under chiral coupling conditions, achieved by matching
the phase difference $\phi $ between coupling points to the propagation
phase of the input field. Focusing on a nonlinear giant cavity coupled
unidirectionally to a waveguide, we demonstrated that in the weak-driving
(few-photon) regime, strongly nonclassical photon statistics can be
deterministically generated---even in the presence of strong dissipation and
weak nonlinearity. This arises from coherent interference between forward-
and backward-propagating components within the nonlocal coupling region and
is tunable via the coupling-point separation, reflecting the intrinsic
non-Markovianity of the system. At the special point $\phi =\pm \mathrm{\pi }%
/2$, interference between coupling points vanishes, rendering the system's
internal dynamics exactly Markovian. Nonetheless, the output fields remain
non-Markovian due to the extended coupling structure. In this regime, we
uncover a nonreciprocal dissipative phase transition in the reflected field,
governed by nonlocal photon correlations. The physical implementations are
presented, along with a discussion of the impact of small deviations from
the chirality condition. Our study can be extended to casese with $N$
coupling legs (see Supplementary Note 4). The findings offer a robust
platform for engineering chiral, non-Markovian multiphoton states in
driven-dissipative settings. This framework is broadly applicable to a
variety of nonlocally coupled quantum systems, including giant cavities
interfaced with atomic ensembles and central spin models coupled to spin
chains, providing a foundation for scalable quantum state control in open
quantum networks.

\section{Methods}

\subsection{Numerical simulations}

The dissipative phase transitions shown in Figs. \ref{fig3} and \ref{fig5}
are obtained by solving the master equation (\ref{2}) and using the
input-output formalism (\ref{3}). the master equation (\ref{2}) is solved in
the superspace, where the density matrix $\rho $ is reshaped as a column
vector $\vec{\rho}$ of dimension $D^{2}$. Here, $D$ is the Hilber space
dimension, with $D-1$ corresponding to the cutoff of the boson number for
mode $b$. In the superspace, the Liouvillian $\mathcal{L}$, defined as%
\begin{equation}
\mathcal{L}\rho \equiv -i\left[ H_{S},\rho \right] +2\gamma \left( 2b\rho
b^{\dagger }-\left\{ b^{\dagger }b,\rho \right\} \right) ,
\end{equation}%
is represented as a $D^{2}\times D^{2}$ matrix%
\begin{eqnarray}
\mathcal{L}_{M} &=&-i\left( H_{S}\otimes I_{D}-I_{D}\otimes \left(
H_{S}\right) ^{T}\right) +  \notag \\
&&2\gamma \left( 2b\otimes \left( b^{\dagger }\right) ^{T}-b^{\dagger
}b\otimes I_{D}-I_{D}\otimes b^{\dagger }b\right) ,
\end{eqnarray}%
where $I_{D}$ is the $D$-dimensional identity matrix, and $\left( \bullet
\right) ^{T}$ denotes the matrix transpose.

The steady state solution $\vec{\rho}_{ss}$ of the master equation can be
obtained by diagonalizing the Liouvillian matrix $\mathcal{L}_{M}$ and
identifying the zero-eigenvalue state. The steady state must satisfy the
normalization condition $u_{0}\vec{\rho}_{ss}=1$, which follows from Tr$%
\left[ \rho \right] =1$, where $u_{0}$ is generated by reshaping the $D$%
-dimensional identity matrix $I_{D}$ into a $1\times D^{2}$ row vector.
Alternatively, $\vec{\rho}_{ss}$ can be found from the equation $\mathcal{L}%
_{M}\vec{\rho}_{ss}=0$. Note that $\mathcal{L}_{M}$ is not reversible (its
rank is less than $D^{2})$, and that is why it has zero eigenvalues.
Physically, this reflects Tr$\left[ \rho \right] =1$, which implies Tr$\left[
\mathcal{L}\rho \right] =0$. To incorporate the normalization condition Tr$%
\left[ \rho \right] =1$, one replaces the row corresponding to the
diagonal-element of $\vec{\rho}$ ($1,D+1,2D+2,...,D^{2}$)with $u_{0}$.
Without loss of generality, we replace the last row of $\mathcal{L}_{M}$
with $u_{0},$ forming a modified matrix $\mathcal{L}_{M}^{\prime }$. The
steady state is then given by $\vec{\rho}_{ss}=\left( \mathcal{L}%
_{M}^{\prime }\right) ^{-1}v_{0}$, where $v_{0}$ is a $D^{2}\times 1$ column
vector with $1$ in the last element and zeros elsewhere.

Once the steady state $\vec{\rho}_{ss}$ is obtained, the long-term-average of
any system operator can be calculated. For example, the mean value photon
number value is given by%
\begin{equation}
\langle b^{\dag }b\rangle _{S}=u_{0}\left( b^{\dag }b\otimes I_{D}\right)
\vec{\rho}_{ss}.
\end{equation}%
Multi-time correlations function can be evaluated using the quantum
regression theorem \cite{Gardiner2004,Walls2008}. For instance, the two-time
correlation function in Eq. (\ref{7}) can be computed as%
\begin{equation}
\left\langle b^{\dag }\left( d\right) b\left( 0\right) \right\rangle
=u_{0}\left( b^{\dag }\otimes I_{D}\right) e^{\mathcal{L}_{M}d}\left(
b\otimes I_{D}\right) \vec{\rho}_{ss}.
\end{equation}

\section{Data availability}

Data that support the findings of this study are available from the
corresponding author upon reasonable request.

\section{Code availability}

The codes used in this study are available from the corresponding author
upon reasonable request.

\section{Acknowledgements}

\acknowledgments The author would like to thank Carlos Navarrete-Benlloch
for fruitful discussions on the physical implementation.


\section{References}

\bibliography{arXiv.bbl}

\begin{thebibliography}{64}
\providecommand{\natexlab}[1]{#1}
\providecommand{\url}[1]{\texttt{#1}}
\expandafter\ifx\csname urlstyle\endcsname\relax
  \providecommand{\doi}[1]{doi: #1}\else
  \providecommand{\doi}{doi: \begingroup \urlstyle{rm}\Url}\fi

\bibitem[Milonni(2013)]{milonni2013quantum}
Peter~W Milonni.
\newblock \emph{The quantum vacuum: an introduction to quantum
  electrodynamics}.
\newblock Academic press, 2013.

\bibitem[Kimble(1998)]{kimble1998strong}
H~Jeff Kimble.
\newblock Strong interactions of single atoms and photons in cavity qed.
\newblock \emph{Physica Scripta}, 1998\penalty0 (T76):\penalty0 127, 1998.

\bibitem[Walls and Milburn(2008)]{Walls2008}
D.~F. Walls and G.~J. Milburn.
\newblock \emph{Quantum Optics}.
\newblock SpringerLink: Springer e-Books. Springer Berlin, 2008.
\newblock ISBN 9783540285731.
\newblock URL \url{https://books.google.com.sg/books?id=LiWsc3Nlf0kC}.

\bibitem[Haroche and Raimond(2006)]{haroche2006exploring}
Serge Haroche and J-M Raimond.
\newblock \emph{Exploring the quantum: atoms, cavities, and photons}.
\newblock Oxford university press, 2006.

\bibitem[Gardiner(1993)]{PhysRevLett.70.2269}
C.~W. Gardiner.
\newblock Driving a quantum system with the output field from another driven
  quantum system.
\newblock \emph{Phys. Rev. Lett.}, 70:\penalty0 2269--2272, Apr 1993.
\newblock \doi{10.1103/PhysRevLett.70.2269}.
\newblock URL \url{https://link.aps.org/doi/10.1103/PhysRevLett.70.2269}.

\bibitem[Carmichael(1993)]{PhysRevLett.70.2273}
H.~J. Carmichael.
\newblock Quantum trajectory theory for cascaded open systems.
\newblock \emph{Phys. Rev. Lett.}, 70:\penalty0 2273--2276, Apr 1993.
\newblock \doi{10.1103/PhysRevLett.70.2273}.
\newblock URL \url{https://link.aps.org/doi/10.1103/PhysRevLett.70.2273}.

\bibitem[Soro and Kockum(2022)]{PhysRevA.105.023712}
Ariadna Soro and Anton~Frisk Kockum.
\newblock Chiral quantum optics with giant atoms.
\newblock \emph{Phys. Rev. A}, 105:\penalty0 023712, Feb 2022.
\newblock \doi{10.1103/PhysRevA.105.023712}.
\newblock URL \url{https://link.aps.org/doi/10.1103/PhysRevA.105.023712}.

\bibitem[Lodahl et~al.(2017)Lodahl, Mahmoodian, Stobbe, Rauschenbeutel,
  Schneeweiss, Volz, Pichler, and Zoller]{lodahl2017chiral}
Peter Lodahl, Sahand Mahmoodian, S{\o}ren Stobbe, Arno Rauschenbeutel, Philipp
  Schneeweiss, J{\"u}rgen Volz, Hannes Pichler, and Peter Zoller.
\newblock Chiral quantum optics.
\newblock \emph{Nature}, 541\penalty0 (7638):\penalty0 473--480, 2017.

\bibitem[Roushan et~al.(2017)Roushan, Neill, Megrant, Chen, Babbush, Barends,
  Campbell, Chen, Chiaro, Dunsworth, et~al.]{roushan2017chiral}
Pedram Roushan, Charles Neill, Anthony Megrant, Yu~Chen, Ryan Babbush, Rami
  Barends, Brooks Campbell, Zijun Chen, Ben Chiaro, Andrew Dunsworth, et~al.
\newblock Chiral ground-state currents of interacting photons in a synthetic
  magnetic field.
\newblock \emph{Nature Physics}, 13\penalty0 (2):\penalty0 146--151, 2017.

\bibitem[Kannan et~al.(2023)Kannan, Almanakly, Sung, Di~Paolo, Rower,
  Braum{\"u}ller, Melville, Niedzielski, Karamlou, Serniak,
  et~al.]{kannan2023demand}
Bharath Kannan, Aziza Almanakly, Youngkyu Sung, Agustin Di~Paolo, David~A
  Rower, Jochen Braum{\"u}ller, Alexander Melville, Bethany~M Niedzielski, Amir
  Karamlou, Kyle Serniak, et~al.
\newblock On-demand directional microwave photon emission using waveguide
  quantum electrodynamics.
\newblock \emph{Nature Physics}, 19\penalty0 (3):\penalty0 394--400, 2023.

\bibitem[Stannigel et~al.(2012)Stannigel, Rabl, and Zoller]{Stannigel_2012}
K~Stannigel, P~Rabl, and P~Zoller.
\newblock Driven-dissipative preparation of entangled states in cascaded
  quantum-optical networks.
\newblock \emph{New Journal of Physics}, 14\penalty0 (6):\penalty0 063014, jun
  2012.
\newblock \doi{10.1088/1367-2630/14/6/063014}.
\newblock URL \url{https://dx.doi.org/10.1088/1367-2630/14/6/063014}.

\bibitem[Ramos et~al.(2014)Ramos, Pichler, Daley, and
  Zoller]{PhysRevLett.113.237203}
Tom\'as Ramos, Hannes Pichler, Andrew~J. Daley, and Peter Zoller.
\newblock Quantum spin dimers from chiral dissipation in cold-atom chains.
\newblock \emph{Phys. Rev. Lett.}, 113:\penalty0 237203, Dec 2014.
\newblock \doi{10.1103/PhysRevLett.113.237203}.
\newblock URL \url{https://link.aps.org/doi/10.1103/PhysRevLett.113.237203}.

\bibitem[Cirac et~al.(1997)Cirac, Zoller, Kimble, and
  Mabuchi]{PhysRevLett.78.3221}
J.~I. Cirac, P.~Zoller, H.~J. Kimble, and H.~Mabuchi.
\newblock Quantum state transfer and entanglement distribution among distant
  nodes in a quantum network.
\newblock \emph{Phys. Rev. Lett.}, 78:\penalty0 3221--3224, Apr 1997.
\newblock \doi{10.1103/PhysRevLett.78.3221}.
\newblock URL \url{https://link.aps.org/doi/10.1103/PhysRevLett.78.3221}.

\bibitem[Kimble(2008)]{Kimble2008}
H.~J. Kimble.
\newblock The quantum internet.
\newblock \emph{Nature}, 453:\penalty0 1023--1030, June 2008.
\newblock \doi{10.1038/nature07127}.
\newblock URL \url{https://doi.org/10.1038/nature07127}.

\bibitem[Petersen et~al.(2014)Petersen, Volz, and Rauschenbeutel]{Petersen2014}
Jan Petersen, Jürgen Volz, and Arno Rauschenbeutel.
\newblock Chiral nanophotonic waveguide interface based on spin-orbit
  interaction of light.
\newblock \emph{Science}, 346\penalty0 (6205):\penalty0 67--71, 2014.
\newblock \doi{10.1126/science.1257671}.
\newblock URL \url{https://www.science.org/doi/abs/10.1126/science.1257671}.

\bibitem[Sayrin et~al.(2015)Sayrin, Junge, Mitsch, Albrecht, O'Shea,
  Schneeweiss, Volz, and Rauschenbeutel]{PhysRevX.5.041036}
Cl\'ement Sayrin, Christian Junge, Rudolf Mitsch, Bernhard Albrecht, Danny
  O'Shea, Philipp Schneeweiss, J\"urgen Volz, and Arno Rauschenbeutel.
\newblock Nanophotonic optical isolator controlled by the internal state of
  cold atoms.
\newblock \emph{Phys. Rev. X}, 5:\penalty0 041036, Dec 2015.
\newblock \doi{10.1103/PhysRevX.5.041036}.
\newblock URL \url{https://link.aps.org/doi/10.1103/PhysRevX.5.041036}.

\bibitem[Scheucher et~al.(2016)Scheucher, Hilico, Will, Volz, and
  Rauschenbeutel]{Scheucher2016}
Michael Scheucher, Adèle Hilico, Elisa Will, Jürgen Volz, and Arno
  Rauschenbeutel.
\newblock Quantum optical circulator controlled by a single chirally coupled
  atom.
\newblock \emph{Science}, 354\penalty0 (6319):\penalty0 1577--1580, 2016.
\newblock \doi{10.1126/science.aaj2118}.
\newblock URL \url{https://www.science.org/doi/abs/10.1126/science.aaj2118}.

\bibitem[Frisk~Kockum et~al.(2014)Frisk~Kockum, Delsing, and
  Johansson]{PhysRevA.90.013837}
Anton Frisk~Kockum, Per Delsing, and G\"oran Johansson.
\newblock Designing frequency-dependent relaxation rates and lamb shifts for a
  giant artificial atom.
\newblock \emph{Phys. Rev. A}, 90:\penalty0 013837, Jul 2014.
\newblock \doi{10.1103/PhysRevA.90.013837}.
\newblock URL \url{https://link.aps.org/doi/10.1103/PhysRevA.90.013837}.

\bibitem[Guo et~al.(2017)Guo, Grimsmo, Kockum, Pletyukhov, and
  Johansson]{PhysRevA.95.053821}
Lingzhen Guo, Arne Grimsmo, Anton~Frisk Kockum, Mikhail Pletyukhov, and G\"oran
  Johansson.
\newblock Giant acoustic atom: A single quantum system with a deterministic
  time delay.
\newblock \emph{Phys. Rev. A}, 95:\penalty0 053821, May 2017.
\newblock \doi{10.1103/PhysRevA.95.053821}.
\newblock URL \url{https://link.aps.org/doi/10.1103/PhysRevA.95.053821}.

\bibitem[Kockum et~al.(2018)Kockum, Johansson, and
  Nori]{PhysRevLett.120.140404}
Anton~Frisk Kockum, G\"oran Johansson, and Franco Nori.
\newblock Decoherence-free interaction between giant atoms in waveguide quantum
  electrodynamics.
\newblock \emph{Phys. Rev. Lett.}, 120:\penalty0 140404, Apr 2018.
\newblock \doi{10.1103/PhysRevLett.120.140404}.
\newblock URL \url{https://link.aps.org/doi/10.1103/PhysRevLett.120.140404}.

\bibitem[Kannan et~al.(2020)Kannan, Ruckriegel, Campbell, Frisk~Kockum,
  Braum{\"u}ller, Kim, Kjaergaard, Krantz, Melville, Niedzielski,
  et~al.]{kannan2020waveguide}
Bharath Kannan, Max~J Ruckriegel, Daniel~L Campbell, Anton Frisk~Kockum, Jochen
  Braum{\"u}ller, David~K Kim, Morten Kjaergaard, Philip Krantz, Alexander
  Melville, Bethany~M Niedzielski, et~al.
\newblock Waveguide quantum electrodynamics with superconducting artificial
  giant atoms.
\newblock \emph{Nature}, 583\penalty0 (7818):\penalty0 775--779, 2020.

\bibitem[Cheng et~al.(2022)Cheng, Wang, and Liu]{PhysRevA.106.033522}
Weijun Cheng, Zhihai Wang, and Yu-xi Liu.
\newblock Topology and retardation effect of a giant atom in a topological
  waveguide.
\newblock \emph{Phys. Rev. A}, 106:\penalty0 033522, Sep 2022.
\newblock \doi{10.1103/PhysRevA.106.033522}.
\newblock URL \url{https://link.aps.org/doi/10.1103/PhysRevA.106.033522}.

\bibitem[Noachtar et~al.(2022)Noachtar, Kn\"orzer, and
  Jonsson]{PhysRevA.106.013702}
David~D. Noachtar, Johannes Kn\"orzer, and Robert~H. Jonsson.
\newblock Nonperturbative treatment of giant atoms using chain transformations.
\newblock \emph{Phys. Rev. A}, 106:\penalty0 013702, Jul 2022.
\newblock \doi{10.1103/PhysRevA.106.013702}.
\newblock URL \url{https://link.aps.org/doi/10.1103/PhysRevA.106.013702}.

\bibitem[Terradas-Brians\'o et~al.(2022)Terradas-Brians\'o,
  Gonz\'alez-Guti\'errez, Nori, Mart\'{\i}n-Moreno, and
  Zueco]{PhysRevA.106.063717}
Sergi Terradas-Brians\'o, Carlos~A. Gonz\'alez-Guti\'errez, Franco Nori, Luis
  Mart\'{\i}n-Moreno, and David Zueco.
\newblock Ultrastrong waveguide qed with giant atoms.
\newblock \emph{Phys. Rev. A}, 106:\penalty0 063717, Dec 2022.
\newblock \doi{10.1103/PhysRevA.106.063717}.
\newblock URL \url{https://link.aps.org/doi/10.1103/PhysRevA.106.063717}.

\bibitem[Du et~al.(2021)Du, Chen, and Li]{PhysRevResearch.3.043226}
Lei Du, Yao-Tong Chen, and Yong Li.
\newblock Nonreciprocal frequency conversion with chiral
  $\mathrm{\ensuremath{\Lambda}}$-type atoms.
\newblock \emph{Phys. Rev. Res.}, 3:\penalty0 043226, Dec 2021.
\newblock \doi{10.1103/PhysRevResearch.3.043226}.
\newblock URL \url{https://link.aps.org/doi/10.1103/PhysRevResearch.3.043226}.

\bibitem[Joshi et~al.(2023)Joshi, Yang, and Mirhosseini]{PhysRevX.13.021039}
Chaitali Joshi, Frank Yang, and Mohammad Mirhosseini.
\newblock Resonance fluorescence of a chiral artificial atom.
\newblock \emph{Phys. Rev. X}, 13:\penalty0 021039, Jun 2023.
\newblock \doi{10.1103/PhysRevX.13.021039}.
\newblock URL \url{https://link.aps.org/doi/10.1103/PhysRevX.13.021039}.

\bibitem[Gu et~al.(2017)Gu, Kockum, Miranowicz, xi~Liu, and Nori]{Gu2017}
Xiu Gu, Anton~Frisk Kockum, Adam Miranowicz, Yu~xi~Liu, and Franco Nori.
\newblock Microwave photonics with superconducting quantum circuits.
\newblock \emph{Physics Reports}, 718-719:\penalty0 1--102, 2017.
\newblock ISSN 0370-1573.
\newblock \doi{https://doi.org/10.1016/j.physrep.2017.10.002}.
\newblock URL
  \url{https://www.sciencedirect.com/science/article/pii/S0370157317303290}.

\bibitem[{Guimond} et~al.(2020){Guimond}, {Vermersch}, {Juan}, {Sharafiev},
  {Kirchmair}, and {Zoller}]{2020npjQI...6...32G}
P.~O. {Guimond}, B.~{Vermersch}, M.~L. {Juan}, A.~{Sharafiev}, G.~{Kirchmair},
  and P.~{Zoller}.
\newblock {A unidirectional on-chip photonic interface for superconducting
  circuits}.
\newblock \emph{npj Quantum Information}, 6:\penalty0 32, March 2020.
\newblock \doi{10.1038/s41534-020-0261-9}.

\bibitem[Wang et~al.(2021)Wang, Liu, Kockum, Li, and
  Nori]{PhysRevLett.126.043602}
Xin Wang, Tao Liu, Anton~Frisk Kockum, Hong-Rong Li, and Franco Nori.
\newblock Tunable chiral bound states with giant atoms.
\newblock \emph{Phys. Rev. Lett.}, 126:\penalty0 043602, Jan 2021.
\newblock \doi{10.1103/PhysRevLett.126.043602}.
\newblock URL \url{https://link.aps.org/doi/10.1103/PhysRevLett.126.043602}.

\bibitem[Chen et~al.(2022)Chen, Du, Guo, Wang, Zhang, Li, and
  Wu]{chen2022nonreciprocal}
Yao-Tong Chen, Lei Du, Lingzhen Guo, Zhihai Wang, Yan Zhang, Yong Li, and
  Jin-Hui Wu.
\newblock Nonreciprocal and chiral single-photon scattering for giant atoms.
\newblock \emph{Communications Physics}, 5\penalty0 (1):\penalty0 215, 2022.

\bibitem[Wang and Li(2022)]{Wang_2022}
Xin Wang and Hong-Rong Li.
\newblock Chiral quantum network with giant atoms.
\newblock \emph{Quantum Science and Technology}, 7\penalty0 (3):\penalty0
  035007, may 2022.
\newblock \doi{10.1088/2058-9565/ac6a04}.
\newblock URL \url{https://dx.doi.org/10.1088/2058-9565/ac6a04}.

\bibitem[Roccati and Cilluffo(2024)]{PhysRevLett.133.063603}
Federico Roccati and Dario Cilluffo.
\newblock Controlling markovianity with chiral giant atoms.
\newblock \emph{Phys. Rev. Lett.}, 133:\penalty0 063603, Aug 2024.
\newblock \doi{10.1103/PhysRevLett.133.063603}.
\newblock URL \url{https://link.aps.org/doi/10.1103/PhysRevLett.133.063603}.

\bibitem[Shi et~al.(2015)Shi, Chang, and Cirac]{PhysRevA.92.053834}
Tao Shi, Darrick~E. Chang, and J.~Ignacio Cirac.
\newblock Multiphoton-scattering theory and generalized master equations.
\newblock \emph{Phys. Rev. A}, 92:\penalty0 053834, Nov 2015.
\newblock \doi{10.1103/PhysRevA.92.053834}.
\newblock URL \url{https://link.aps.org/doi/10.1103/PhysRevA.92.053834}.

\bibitem[Gu et~al.(2024)Gu, Li, Tian, Yi, and Li]{PhysRevA.110.033707}
Wenju Gu, Tao Li, Ye~Tian, Zhen Yi, and Gao-xiang Li.
\newblock Two-photon dynamics in non-markovian waveguide qed with a giant atom.
\newblock \emph{Phys. Rev. A}, 110:\penalty0 033707, Sep 2024.
\newblock \doi{10.1103/PhysRevA.110.033707}.
\newblock URL \url{https://link.aps.org/doi/10.1103/PhysRevA.110.033707}.

\bibitem[Chang et~al.(2016)Chang, Gonz\'alez-Tudela, S\'anchez Mu\~noz,
  Navarrete-Benlloch, and Shi]{PhysRevLett.117.203602}
Yue Chang, Alejandro Gonz\'alez-Tudela, Carlos S\'anchez Mu\~noz, Carlos
  Navarrete-Benlloch, and Tao Shi.
\newblock Deterministic down-converter and continuous photon-pair source within
  the bad-cavity limit.
\newblock \emph{Phys. Rev. Lett.}, 117:\penalty0 203602, Nov 2016.
\newblock \doi{10.1103/PhysRevLett.117.203602}.
\newblock URL \url{https://link.aps.org/doi/10.1103/PhysRevLett.117.203602}.

\bibitem[Kong et~al.(2025)Kong, Navarrete-Benlloch, and
  Chang]{PhysRevResearch.7.L022024}
Xiangjin Kong, Carlos Navarrete-Benlloch, and Yue Chang.
\newblock Accessing strongly coupled systems without compromising them.
\newblock \emph{Phys. Rev. Res.}, 7:\penalty0 L022024, Apr 2025.
\newblock \doi{10.1103/PhysRevResearch.7.L022024}.
\newblock URL \url{https://link.aps.org/doi/10.1103/PhysRevResearch.7.L022024}.

\bibitem[Yang and Chang(2024)]{PhysRevA.110.023722}
Li-Ping Yang and Yue Chang.
\newblock Quantum beam splitter as a controller of higher-order quantum
  coherence.
\newblock \emph{Phys. Rev. A}, 110:\penalty0 023722, Aug 2024.
\newblock \doi{10.1103/PhysRevA.110.023722}.
\newblock URL \url{https://link.aps.org/doi/10.1103/PhysRevA.110.023722}.

\bibitem[Gardiner and Zoller(2004)]{Gardiner2004}
C.~Gardiner and P.~Zoller.
\newblock \emph{Quantum Noise: A Handbook of Markovian and Non-Markovian
  Quantum Stochastic Methods with Applications to Quantum Optics}.
\newblock Springer Series in Synergetics. Springer,Berlin, Heidelberg, 2004.
\newblock ISBN 9783540223016.
\newblock URL \url{https://books.google.com.sg/books?id=a\_xsT8oGhdgC}.

\bibitem[Houck et~al.(2012)Houck, T{\"u}reci, and Koch]{houck2012chip}
Andrew~A Houck, Hakan~E T{\"u}reci, and Jens Koch.
\newblock On-chip quantum simulation with superconducting circuits.
\newblock \emph{Nature Physics}, 8\penalty0 (4):\penalty0 292--299, 2012.

\bibitem[Kessler et~al.(2012)Kessler, Giedke, Imamoglu, Yelin, Lukin, and
  Cirac]{PhysRevA.86.012116}
E.~M. Kessler, G.~Giedke, A.~Imamoglu, S.~F. Yelin, M.~D. Lukin, and J.~I.
  Cirac.
\newblock Dissipative phase transition in a central spin system.
\newblock \emph{Phys. Rev. A}, 86:\penalty0 012116, Jul 2012.
\newblock \doi{10.1103/PhysRevA.86.012116}.
\newblock URL \url{https://link.aps.org/doi/10.1103/PhysRevA.86.012116}.

\bibitem[Diehl et~al.(2008)Diehl, Micheli, Kantian, Kraus, B{\"u}chler, and
  Zoller]{diehl2008quantum}
Sebastian Diehl, A~Micheli, Adrian Kantian, B~Kraus, HP~B{\"u}chler, and Peter
  Zoller.
\newblock Quantum states and phases in driven open quantum systems with cold
  atoms.
\newblock \emph{Nature Physics}, 4\penalty0 (11):\penalty0 878--883, 2008.

\bibitem[Carmichael(2015)]{PhysRevX.5.031028}
H.~J. Carmichael.
\newblock Breakdown of photon blockade: A dissipative quantum phase transition
  in zero dimensions.
\newblock \emph{Phys. Rev. X}, 5:\penalty0 031028, Sep 2015.
\newblock \doi{10.1103/PhysRevX.5.031028}.
\newblock URL \url{https://link.aps.org/doi/10.1103/PhysRevX.5.031028}.

\bibitem[Fitzpatrick et~al.(2017)Fitzpatrick, Sundaresan, Li, Koch, and
  Houck]{PhysRevX.7.011016}
Mattias Fitzpatrick, Neereja~M. Sundaresan, Andy C.~Y. Li, Jens Koch, and
  Andrew~A. Houck.
\newblock Observation of a dissipative phase transition in a one-dimensional
  circuit qed lattice.
\newblock \emph{Phys. Rev. X}, 7:\penalty0 011016, Feb 2017.
\newblock \doi{10.1103/PhysRevX.7.011016}.
\newblock URL \url{https://link.aps.org/doi/10.1103/PhysRevX.7.011016}.

\bibitem[Aspelmeyer et~al.(2014)Aspelmeyer, Kippenberg, and
  Marquardt]{RevModPhys.86.1391}
Markus Aspelmeyer, Tobias~J. Kippenberg, and Florian Marquardt.
\newblock Cavity optomechanics.
\newblock \emph{Rev. Mod. Phys.}, 86:\penalty0 1391--1452, Dec 2014.
\newblock \doi{10.1103/RevModPhys.86.1391}.
\newblock URL \url{https://link.aps.org/doi/10.1103/RevModPhys.86.1391}.

\bibitem[Benito et~al.(2016)Benito, S\'anchez Mu\~noz, and
  Navarrete-Benlloch]{PhysRevA.93.023846}
M\'onica Benito, Carlos S\'anchez Mu\~noz, and Carlos Navarrete-Benlloch.
\newblock Degenerate parametric oscillation in quantum membrane optomechanics.
\newblock \emph{Phys. Rev. A}, 93:\penalty0 023846, Feb 2016.
\newblock \doi{10.1103/PhysRevA.93.023846}.
\newblock URL \url{https://link.aps.org/doi/10.1103/PhysRevA.93.023846}.

\bibitem[Minganti et~al.(2018)Minganti, Biella, Bartolo, and
  Ciuti]{PhysRevA.98.042118}
Fabrizio Minganti, Alberto Biella, Nicola Bartolo, and Cristiano Ciuti.
\newblock Spectral theory of liouvillians for dissipative phase transitions.
\newblock \emph{Phys. Rev. A}, 98:\penalty0 042118, Oct 2018.
\newblock \doi{10.1103/PhysRevA.98.042118}.
\newblock URL \url{https://link.aps.org/doi/10.1103/PhysRevA.98.042118}.

\bibitem[Imamo\ifmmode~\bar{g}\else \={g}\fi{}lu
  et~al.(1997)Imamo\ifmmode~\bar{g}\else \={g}\fi{}lu, Schmidt, Woods, and
  Deutsch]{PhysRevLett.79.1467}
A.~Imamo\ifmmode~\bar{g}\else \={g}\fi{}lu, H.~Schmidt, G.~Woods, and
  M.~Deutsch.
\newblock Strongly interacting photons in a nonlinear cavity.
\newblock \emph{Phys. Rev. Lett.}, 79:\penalty0 1467--1470, Aug 1997.
\newblock \doi{10.1103/PhysRevLett.79.1467}.
\newblock URL \url{https://link.aps.org/doi/10.1103/PhysRevLett.79.1467}.

\bibitem[Rebi\ifmmode~\acute{c}\else \'{c}\fi{}
  et~al.(2009)Rebi\ifmmode~\acute{c}\else \'{c}\fi{}, Twamley, and
  Milburn]{PhysRevLett.103.150503}
Stojan Rebi\ifmmode~\acute{c}\else \'{c}\fi{}, Jason Twamley, and Gerard~J.
  Milburn.
\newblock Giant kerr nonlinearities in circuit quantum electrodynamics.
\newblock \emph{Phys. Rev. Lett.}, 103:\penalty0 150503, Oct 2009.
\newblock \doi{10.1103/PhysRevLett.103.150503}.
\newblock URL \url{https://link.aps.org/doi/10.1103/PhysRevLett.103.150503}.

\bibitem[Blais et~al.(2021)Blais, Grimsmo, Girvin, and
  Wallraff]{RevModPhys.93.025005}
Alexandre Blais, Arne~L. Grimsmo, S.~M. Girvin, and Andreas Wallraff.
\newblock Circuit quantum electrodynamics.
\newblock \emph{Rev. Mod. Phys.}, 93:\penalty0 025005, May 2021.
\newblock \doi{10.1103/RevModPhys.93.025005}.
\newblock URL \url{https://link.aps.org/doi/10.1103/RevModPhys.93.025005}.

\bibitem[Chang et~al.(2011)Chang, Gong, and Sun]{PhysRevA.83.013825}
Yue Chang, Z.~R. Gong, and C.~P. Sun.
\newblock Multiatomic mirror for perfect reflection of single photons in a wide
  band of frequency.
\newblock \emph{Phys. Rev. A}, 83:\penalty0 013825, Jan 2011.
\newblock \doi{10.1103/PhysRevA.83.013825}.
\newblock URL \url{https://link.aps.org/doi/10.1103/PhysRevA.83.013825}.

\bibitem[Shi et~al.(2016)Shi, Wu, Gonz\'alez-Tudela, and
  Cirac]{PhysRevX.6.021027}
Tao Shi, Ying-Hai Wu, A.~Gonz\'alez-Tudela, and J.~I. Cirac.
\newblock Bound states in boson impurity models.
\newblock \emph{Phys. Rev. X}, 6:\penalty0 021027, May 2016.
\newblock \doi{10.1103/PhysRevX.6.021027}.
\newblock URL \url{https://link.aps.org/doi/10.1103/PhysRevX.6.021027}.

\bibitem[Shi et~al.(2018)Shi, Chang, and
  Garc\'{\i}a-Ripoll]{PhysRevLett.120.153602}
Tao Shi, Yue Chang, and Juan~Jos\'e Garc\'{\i}a-Ripoll.
\newblock Ultrastrong coupling few-photon scattering theory.
\newblock \emph{Phys. Rev. Lett.}, 120:\penalty0 153602, Apr 2018.
\newblock \doi{10.1103/PhysRevLett.120.153602}.
\newblock URL \url{https://link.aps.org/doi/10.1103/PhysRevLett.120.153602}.

\bibitem[Sakurai and Napolitano(2020)]{Sakurai_Napolitano_2020}
J.~J. Sakurai and Jim Napolitano.
\newblock \emph{Modern Quantum Mechanics}.
\newblock Cambridge University Press, 3 edition, 2020.

\bibitem[Fetter and Walecka(2012)]{fetter2012quantum}
Alexander~L. Fetter and John~Dirk Walecka.
\newblock \emph{Quantum Theory of Many-Particle Systems}.
\newblock Dover, Mineola, New York, 2012.

\bibitem[Shen and Fan(2007)]{PhysRevLett.98.153003}
Jung-Tsung Shen and Shanhui Fan.
\newblock Strongly correlated two-photon transport in a one-dimensional
  waveguide coupled to a two-level system.
\newblock \emph{Phys. Rev. Lett.}, 98:\penalty0 153003, Apr 2007.
\newblock \doi{10.1103/PhysRevLett.98.153003}.
\newblock URL \url{https://link.aps.org/doi/10.1103/PhysRevLett.98.153003}.

\bibitem[Shi and Sun(2009)]{PhysRevB.79.205111}
T.~Shi and C.~P. Sun.
\newblock Lehmann-symanzik-zimmermann reduction approach to multiphoton
  scattering in coupled-resonator arrays.
\newblock \emph{Phys. Rev. B}, 79:\penalty0 205111, May 2009.
\newblock \doi{10.1103/PhysRevB.79.205111}.
\newblock URL \url{https://link.aps.org/doi/10.1103/PhysRevB.79.205111}.

\bibitem[Shi et~al.(2011)Shi, Fan, and Sun]{PhysRevA.84.063803}
T.~Shi, Shanhui Fan, and C.~P. Sun.
\newblock Two-photon transport in a waveguide coupled to a cavity in a
  two-level system.
\newblock \emph{Phys. Rev. A}, 84:\penalty0 063803, Dec 2011.
\newblock \doi{10.1103/PhysRevA.84.063803}.
\newblock URL \url{https://link.aps.org/doi/10.1103/PhysRevA.84.063803}.

\bibitem[Drummond and Walls(1980)]{drummond1980quantum}
PD~Drummond and DF~Walls.
\newblock Quantum theory of optical bistability. i. nonlinear polarisability
  model.
\newblock \emph{Journal of Physics A: Mathematical and General}, 13\penalty0
  (2):\penalty0 725, 1980.

\bibitem[Drummond and Gardiner(1980)]{drummond1980generalised}
Peter~D Drummond and Crispin~W Gardiner.
\newblock Generalised p-representations in quantum optics.
\newblock \emph{Journal of Physics A: Mathematical and General}, 13\penalty0
  (7):\penalty0 2353, 1980.

\bibitem[Bartolo et~al.(2016)Bartolo, Minganti, Casteels, and
  Ciuti]{PhysRevA.94.033841}
Nicola Bartolo, Fabrizio Minganti, Wim Casteels, and Cristiano Ciuti.
\newblock Exact steady state of a kerr resonator with one- and two-photon
  driving and dissipation: Controllable wigner-function multimodality and
  dissipative phase transitions.
\newblock \emph{Phys. Rev. A}, 94:\penalty0 033841, Sep 2016.
\newblock \doi{10.1103/PhysRevA.94.033841}.
\newblock URL \url{https://link.aps.org/doi/10.1103/PhysRevA.94.033841}.

\bibitem[Roberts and Clerk(2020)]{PhysRevX.10.021022}
David Roberts and Aashish~A. Clerk.
\newblock Driven-dissipative quantum kerr resonators: New exact solutions,
  photon blockade and quantum bistability.
\newblock \emph{Phys. Rev. X}, 10:\penalty0 021022, Apr 2020.
\newblock \doi{10.1103/PhysRevX.10.021022}.
\newblock URL \url{https://link.aps.org/doi/10.1103/PhysRevX.10.021022}.

\bibitem[Yin et~al.(2013)Yin, Chen, Sank, O'Malley, White, Barends, Kelly,
  Lucero, Mariantoni, Megrant, Neill, Vainsencher, Wenner, Korotkov, Cleland,
  and Martinis]{PhysRevLett.110.107001}
Yi~Yin, Yu~Chen, Daniel Sank, P.~J.~J. O'Malley, T.~C. White, R.~Barends,
  J.~Kelly, Erik Lucero, Matteo Mariantoni, A.~Megrant, C.~Neill,
  A.~Vainsencher, J.~Wenner, Alexander~N. Korotkov, A.~N. Cleland, and John~M.
  Martinis.
\newblock Catch and release of microwave photon states.
\newblock \emph{Phys. Rev. Lett.}, 110:\penalty0 107001, Mar 2013.
\newblock \doi{10.1103/PhysRevLett.110.107001}.
\newblock URL \url{https://link.aps.org/doi/10.1103/PhysRevLett.110.107001}.

\bibitem[Chen et~al.(2014)Chen, Neill, Roushan, Leung, Fang, Barends, Kelly,
  Campbell, Chen, Chiaro, Dunsworth, Jeffrey, Megrant, Mutus, O'Malley,
  Quintana, Sank, Vainsencher, Wenner, White, Geller, Cleland, and
  Martinis]{PhysRevLett.113.220502}
Yu~Chen, C.~Neill, P.~Roushan, N.~Leung, M.~Fang, R.~Barends, J.~Kelly,
  B.~Campbell, Z.~Chen, B.~Chiaro, A.~Dunsworth, E.~Jeffrey, A.~Megrant, J.~Y.
  Mutus, P.~J.~J. O'Malley, C.~M. Quintana, D.~Sank, A.~Vainsencher, J.~Wenner,
  T.~C. White, Michael~R. Geller, A.~N. Cleland, and John~M. Martinis.
\newblock Qubit architecture with high coherence and fast tunable coupling.
\newblock \emph{Phys. Rev. Lett.}, 113:\penalty0 220502, Nov 2014.
\newblock \doi{10.1103/PhysRevLett.113.220502}.
\newblock URL \url{https://link.aps.org/doi/10.1103/PhysRevLett.113.220502}.

\bibitem[Peropadre et~al.(2013)Peropadre, Zueco, Wulschner, Deppe, Marx, Gross,
  and Garc\'{\i}a-Ripoll]{PhysRevB.87.134504}
Borja Peropadre, David Zueco, Friedrich Wulschner, Frank Deppe, Achim Marx,
  Rudolf Gross, and Juan~Jos\'e Garc\'{\i}a-Ripoll.
\newblock Tunable coupling engineering between superconducting resonators: From
  sidebands to effective gauge fields.
\newblock \emph{Phys. Rev. B}, 87:\penalty0 134504, Apr 2013.
\newblock \doi{10.1103/PhysRevB.87.134504}.
\newblock URL \url{https://link.aps.org/doi/10.1103/PhysRevB.87.134504}.

\end{thebibliography}

\section{Contributions}

The sole author was responsible for all aspects of this work, including
conceptualization, calculations, interpretation, and manuscript preparation.

\section{Competing interests}

The authors declare no competing interests.

\newpage \widetext

\begin{center}
\textbf{\large Supplemental information for ``Non-Markovian Multiphoton Chiral Dynamics with Giant Systems ''}
\end{center}

\setcounter{equation}{0} \setcounter{figure}{0} 
\makeatletter

\renewcommand{\thefigure}{SM\arabic{figure}} \renewcommand{\thesection}{SM%
\arabic{section}} \renewcommand{\theequation}{SM\arabic{equation}}

\setcounter{equation}{0} \setcounter{figure}{0} \setcounter{table}{0} %
\setcounter{page}{1} \setcounter{section}{0} \makeatletter
\renewcommand{\theequation}{S\arabic{equation}} \renewcommand{\thefigure}{S%
\arabic{figure}} \renewcommand{\bibnumfmt}[1]{[S#1]} \renewcommand{%
\citenumfont}[1]{S#1} \renewcommand{\thesection}{S\arabic{section}}%
\setcounter{secnumdepth}{3}

\section*{Supplementary Note1: S-matrix in coherent-state presentation}

We first rewrite the $S$ matrix for
reflection and transmission photons in forms of the two modes $A_{\pm k}$
defined in Eq. (11) as%
\begin{equation}
\left(
\begin{array}{c}
A_{+k} \\
A_{-k}%
\end{array}%
\right) =U_{k}\left(
\begin{array}{c}
a_{L-k} \\
a_{Rk}%
\end{array}%
\right) .
\end{equation}%
Since only the modes $A_{+k}$ couple to the giant system, the transmission
amplitude $\left\langle 0\right\vert a_{Rk_{f}}Sa_{Rk_{i}}^{\dagger
}\left\vert 0\right\rangle $ for a right-moving photon with input momentum $%
k_{i}$ and an output photon with momentum $k_{f}$ is%
\begin{eqnarray}
\left\langle 0\right\vert a_{Rk_{f}}Sa_{Rk_{i}}^{\dagger }\left\vert
0\right\rangle  &=&\left\vert U_{k_{i},22}\right\vert ^{2}\delta \left(
k_{i}-k_{f}\right) +\left\vert U_{k_{i},12}\right\vert ^{2}\left\langle
0\right\vert A_{k_{f}+}SA_{k_{i}+}^{\dagger }\left\vert 0\right\rangle
\notag \\
&=&\left\vert U_{k_{i},22}\right\vert ^{2}\delta \left( k_{i}-k_{f}\right)
+\left\vert U_{k_{i},12}\right\vert ^{2}\left\langle 0\right\vert
A_{k_{f}+}e^{-i2HT}A_{k_{i}+}^{\dagger }\left\vert 0\right\rangle e^{i\left(
k_{i}+k_{f}\right) T},
\end{eqnarray}%
where $T\rightarrow +\infty $. Similarly, the reflection amplitude%
\begin{equation}
\left\langle 0\right\vert a_{Lk_{f}}Sa_{Rk_{i}}^{\dagger }\left\vert
0\right\rangle =U_{-k_{f},21}^{\ast }U_{k_{i},22}\delta \left(
k_{i}+k_{f}\right) +U_{-k_{f},11}^{\ast }U_{k_{i},12}\left\langle
0\right\vert A_{-k_{f}+}e^{-2HT}A_{k_{i}+}^{\dagger }\left\vert
0\right\rangle e^{ik_{i}T}e^{-ik_{f}T}.
\end{equation}%
For two photons, the amplitude for right-moving incident photons with
momenta $k_{1}$ and $k_{2}$, and transmitted photons with momenta $p_{1}$
and $p_{2}$, is%
\begin{eqnarray}
&&\left\langle 0\right\vert a_{Rp_{1}}a_{Rp_{2}}Sa_{Rk_{2}}^{\dagger
}a_{Rk_{1}}^{\dagger }\left\vert 0\right\rangle   \notag \\
&=&U_{p_{2},22}^{\ast }U_{p_{1},22}^{\ast }U_{k_{2},22}U_{k_{1},22}\left[
\delta \left( k_{1}-p_{1}\right) \delta \left( k_{2}-p_{2}\right) +\left(
k_{1}\longleftrightarrow k_{2}\right) \right]   \notag \\
&&+\{U_{p_{2},22}^{\ast }U_{p_{1},12}^{\ast
}U_{k_{2},22}U_{k_{1},12}\left\langle 0\right\vert
A_{p_{1}+}e^{-i2HT}A_{k_{1}+}^{\dagger }\left\vert 0\right\rangle e^{i\left(
k_{1}+p_{1}\right) T}\delta \left( k_{2}-p_{2}\right)   \notag \\
&&+\left( k_{1}\longleftrightarrow k_{2}\right) +\left(
p_{1}\longleftrightarrow p_{2}\right) +\left( k_{1}\longleftrightarrow
k_{2},p_{1}\longleftrightarrow p_{2}\right) \}  \notag \\
&&+U_{p_{2},12}^{\ast }U_{p_{1},12}^{\ast
}U_{k_{2},12}U_{k_{1},12}\left\langle 0\right\vert
A_{p_{2}+}A_{p_{1}+}e^{-i2HT}A_{k_{2}+}^{\dagger }A_{k_{1}+}^{\dagger
}\left\vert 0\right\rangle e^{i\left( k_{1}+k_{2}+p_{1}+p_{2}\right) T}.
\label{sm4}
\end{eqnarray}%
Therefore, instead of the original left- and right-moving modes $a_{Lk}$ and
$a_{L-k}$, we can obtain the S-matrix by considering only the $A_{k+}$ mode
while the $A_{k-}$ mode is decoupled from the giant system. Here, we have
assumed that in the initial and final states, the giant system is in vacuum,
which is the case we focus on in the weak-driving limit.

To calculate the S-matrix ultilizing the path-integral method, we write it
in the basis of coherent-states. The S-matrix for $N$ input photons with
momenta $k_{i_{1}}k_{i_{2}},...,k_{i_{N}}$\ and $N$ output photons with
momenta $k_{f_{1}}k_{f_{2}},...,k_{f_{N}}$\ envolves the calculation of
\begin{equation}
\left\langle k_{f_{1}}k_{f_{2}}...k_{f_{N}}\right\vert e^{-i2HT}\left\vert
k_{i_{1}}k_{i_{2}}...k_{i_{N}}\right\rangle ,
\end{equation}%
where $\left\vert k_{1}k_{2}...k_{N}\right\rangle =A_{k_{N}+}^{\dagger
}...A_{k_{2}+}^{\dagger }A_{k_{1}+}^{\dagger }\left\vert 0\right\rangle $.
Introducing an unnormalized coherent state \cite{Walls2008}%
\begin{equation}
\left\vert \alpha _{k}\right\rangle =\sum_{n_{k}=0}\frac{\alpha _{k}^{n}}{%
\sqrt{n_{k}!}}\left\vert n_{k}\right\rangle ,
\end{equation}%
for the $k$-th mode $A_{k+}^{\dagger }$, the the evolution term $%
\left\langle k_{f_{1}}k_{f_{2}}...k_{f_{N}}\right\vert e^{-i2HT}\left\vert
k_{i_{1}}k_{i_{2}}...k_{i_{N}}\right\rangle $ can be written as%
\begin{eqnarray}
&&\left\langle k_{f_{1}}k_{f_{2}}...k_{f_{N}}\right\vert e^{-i2HT}\left\vert
k_{i_{1}}k_{i_{2}}...k_{i_{N}}\right\rangle  \notag \\
&=&\lim_{\alpha _{k_{i_{1}}}\rightarrow 0}\partial _{\alpha
_{k_{i_{1}}}}\lim_{\alpha _{k_{i_{2}}}\rightarrow 0}\partial _{\alpha
_{k_{i_{2}}}}...\lim_{\alpha _{k_{i_{N}}}\rightarrow 0}\partial _{\alpha
_{k_{i_{N}}}}\lim_{\alpha _{k_{f_{1}}^{\ast }}\rightarrow 0}\partial
_{\alpha _{k_{f_{1}}^{\ast }}}\lim_{\alpha _{k_{f_{2}}^{\ast }}\rightarrow
0}\partial _{\alpha _{k_{f_{2}}^{\ast }}}...\lim_{\alpha _{k_{f_{N}}^{\ast
}}\rightarrow 0}\partial _{\alpha _{k_{f_{N}}^{\ast }}}\mathcal{A},
\label{sm1}
\end{eqnarray}%
where $\mathcal{A}\equiv \left\langle \alpha _{k_{f_{1}}}\alpha
_{k_{f_{2}}}...\alpha _{k_{f_{N}}}\right\vert e^{-i2HT}\left\vert \alpha
_{k_{i_{1}}}\alpha _{k_{i_{2}}}...\alpha _{k_{i_{N}}}\right\rangle $. Eq. (%
\ref{sm1})\ connects the S-matrix in the fock-state basis to the
coherent-state representaion. Note that the overlap of two reduced coherent
states is%
\begin{equation}
\left\langle \beta \right. \left\vert \alpha \right\rangle =e^{\alpha \beta
^{\ast }}
\end{equation}%
and the completeness property of the reduced coherent states reads%
\begin{equation}
\int \mathrm{d}^{2}\alpha \text{ }\frac{e^{-\left\vert \alpha \right\vert
^{2}}}{\pi }\left\vert \alpha \right\rangle \left\langle \alpha \right\vert
=1.
\end{equation}

The amplitude $\mathcal{A}$ in the path-integral form is \cite%
{PhysRevA.92.053834}%
\begin{equation}
\mathcal{A}=\int D\left[ \beta ,\beta ^{\ast },\alpha _{k},\alpha _{k}^{\ast
}\right] e^{\sum_{k}\alpha _{k}^{\ast }\left( -T\right) \alpha _{k}\left(
-T\right) }\exp \left[ i\int_{-T}^{T}\mathrm{d}t\text{ }\left[ \mathcal{L}%
_{w}\left( t\right) +\mathcal{L}_{c}\left( t\right) \right] \right] ,
\end{equation}%
where%
\begin{equation}
D\left[ \beta ,\beta ^{\ast },\alpha _{k},\alpha _{k}^{\ast }\right]
=\prod\limits_{t}\left[ \frac{d\beta \left( t\right) d\beta ^{\ast }\left(
t\right) }{\pi }\prod\limits_{k}\frac{d\alpha _{k}\left( t\right) d\alpha
_{k}^{\ast }\left( t\right) }{\pi }\right] ,
\end{equation}%
is the measure%
\begin{equation}
\mathcal{L}_{w}\left( t\right) =-i\sum_{k}\dot{\alpha}_{k}^{\ast }\left(
t\right) \alpha _{k}\left( t\right) -\sum_{k}\left[ k\alpha _{k}^{\ast
}\left( t\right) \alpha _{k}\left( t\right) +\tilde{V}_{k}\beta ^{\ast
}\left( t\right) \alpha _{k}\left( t\right) +\tilde{V}_{k}\alpha _{k}^{\ast
}\left( t\right) \beta \left( t\right) \right] ,
\end{equation}%
is the Lagrangian for the waveguide and the emitter-waveguide interaction,
and%
\begin{equation}
\mathcal{L}_{c}\left( t\right) =-i\dot{\beta}^{\ast }\left( t\right) \beta
\left( t\right) -\frac{U}{2}\beta ^{\ast }\left( t\right) \beta ^{\ast
}\left( t\right) \beta \left( t\right) \beta \left( t\right) ,
\end{equation}%
is the Lagrangian for the giant cavity. At the the initial and final
moments, $\beta \left( -T\right) =\beta \left( T\right) =0$, $\left\{ \alpha
_{k}\left( -T\right) \right\} =\left\{ \alpha _{k_{i}}\right\} $, and $%
\left\{ \alpha _{k}\left( T\right) \right\} =\left\{ \alpha _{k_{f}}\right\}
$. Following the saddle-point method \cite{PhysRevA.92.053834}, we first
calculate the classical trajectory given by $\delta _{\alpha _{k}}\mathcal{L}%
_{ph}-\partial _{t}\frac{\delta \mathcal{L}_{ph}}{\delta _{\dot{\alpha}_{k}}}%
=0$ and obtain%
\begin{equation}
\alpha _{k,cl}^{\ast }\left( t\right) =\alpha _{k}^{\ast }\left( T\right)
e^{-ik\left( T-t\right) }-iV_{k}\int_{t}^{T}\beta ^{\ast }\left( t^{\prime
}\right) e^{-ik\left( t^{\prime }-t\right) }\mathrm{d}t^{\prime }.
\end{equation}%
Consequently, the waveguide modes can be integrated out and the amplitude $%
\mathcal{A}$ becomes%
\begin{eqnarray}
\mathcal{A} &=&\int D\left[ \beta ,\beta ^{\ast }\right] \exp \left[
i\int_{-T}^{T}\mathrm{d}t\text{ }\mathcal{L}_{eff}\left( t\right) \right]
\times  \notag \\
&&\exp \left[ \alpha _{k_{f}}^{\ast }\alpha _{k_{i}}e^{-i2k_{i}T}\delta
\left( k_{i}-k_{f}\right) -i\tilde{V}_{k_{i}}\alpha _{k_{i}}\int_{-T}^{T}%
\mathrm{d}t\text{ }\beta ^{\ast }\left( t\right) e^{-ik_{i}\left( t+T\right)
}-i\tilde{V}_{k_{f}}\alpha _{k_{f}}^{\ast }\int_{-T}^{T}\mathrm{d}t\text{ }%
e^{-ik_{f}\left( T-t\right) }\beta \left( t\right) \right] .
\end{eqnarray}%
where%
\begin{equation}
\mathcal{L}_{eff}\left( t\right) =\mathcal{L}_{c}\left( t\right)
+i\int_{t}^{T}\mathrm{d}t^{\prime }\int \mathrm{d}k\text{ }\tilde{V}%
_{k}^{2}e^{-ik\left( t^{\prime }-t\right) }\beta ^{\ast }\left( t^{\prime
}\right) \beta \left( t\right)  \label{sm2}
\end{equation}%
is the effective Lagrangian for the giant system. We will see in the
following that the second term in Eq. (\ref{sm2}) will result in decay and
energy correction to the giant system.

\section*{Supplementary Note 2: Single-photon scattering}

The the S-matrix element $\left\langle 0\right\vert
A_{k_{f}+}SA_{k_{i}+}^{\dagger }\left\vert 0\right\rangle $ for an incident
photon with momentum $k_{i}$ and outgoing one with momentum $k_{f}$ is%
\begin{eqnarray}
&&\left\langle 0\right\vert A_{k_{f}+}SA_{k_{i}+}^{\dagger }\left\vert
0\right\rangle  \notag \\
&=&e^{i\left( k_{i}+k_{f}\right) T}\lim_{\left\{ \alpha _{k}\right\}
\rightarrow 0}\partial _{\alpha _{k_{i}}}\partial _{\alpha _{k_{f}}^{\ast }}%
\mathcal{A}  \notag \\
&=&\int D\left[ \beta ,\beta ^{\ast }\right] \left[ \delta \left(
k_{i}-k_{f}\right) -\tilde{V}_{k_{i}}\tilde{V}_{k_{f}}\int_{-T}^{T}\mathrm{d}%
t\mathrm{d}t^{\prime }e^{-ik_{i}t^{\prime }}e^{ik_{f}t}\beta ^{\ast }\left(
t^{\prime }\right) \beta \left( t\right) \right] \exp \left[ i\int_{-T}^{T}%
\mathrm{d}t\text{ }\mathcal{L}_{eff}\left( t\right) \right] .
\end{eqnarray}%
Denoting the Fourier transform of $\beta \left( t\right) $ as%
\begin{equation}
\beta \left( t\right) =\frac{1}{\sqrt{2\pi }}\int \tilde{\beta}\left( \omega
\right) e^{-i\omega t}\mathrm{d}\omega ,
\end{equation}%
we acquire%
\begin{eqnarray}
&&\left\langle 0\right\vert A_{k_{f}+}SA_{k_{i}+}^{\dagger }\left\vert
0\right\rangle  \notag \\
&=&\delta \left( k_{i}-k_{f}\right) -i2\pi \tilde{V}_{k_{i}}\tilde{V}%
_{k_{f}}\int D\left[ \tilde{\beta},\tilde{\beta}^{\ast }\right] \tilde{\beta}%
^{\ast }\left( k_{i}\right) \tilde{\beta}\left( k_{f}\right) \exp \left[
\int \mathrm{d}\omega \left[ \omega -\Sigma \left( \omega \right) \right]
\tilde{\beta}^{\ast }\left( \omega \right) \tilde{\beta}\left( \omega
\right) \right]  \notag \\
&=&\delta \left( k_{i}-k_{f}\right) -i2\pi \tilde{V}_{k_{i}}^{2}G\left(
k_{i}\right) \delta \left( k_{i}-k_{f}\right) .
\end{eqnarray}%
where the Green's function%
\begin{equation}
G\left( \omega \right) =\frac{1}{\omega -\Sigma \left( \omega \right) }
\end{equation}%
and the self energy%
\begin{equation}
\Sigma \left( \omega \right) \equiv \int \mathrm{d}k\frac{\tilde{V}_{k}^{2}}{%
\omega -k+i0^{+}}=-i2\gamma \left( 1+\cos \phi e^{i\left( k_{0}+\omega
\right) d}\right) .
\end{equation}%
Here, the real part Re$\left[ \Sigma \left( \omega \right) \right] =2\gamma
\cos \phi \sin \left[ \left( k_{0}+\omega \right) d\right] $\ of the self
energy is the frequency correction to the giant system due to the
waveguide-emitter coupling, while the imaginary part -Im$\left[ \Sigma
\left( \omega \right) \right] =2\gamma \left( 1+\cos \phi \cos \left[ \left(
k_{0}+\omega \right) d\right] \right) $ is the waveguide-induced decay rate.
We note that in the Markov approximation, $\Sigma \left( \omega \right) $ is
taken as the frequency-independent self energy $\Sigma \left( 0\right) $,
which is generally valid when $\gamma d\ll 1$.

Consequently, the single-photon transmission amplitude is%
\begin{equation}
\left\langle 0\right\vert a_{Rk_{f}}Sa_{Rk_{i}}^{\dagger }\left\vert
0\right\rangle =\delta \left( k_{i}-k_{f}\right) -i2\gamma \left\vert
e^{i\phi }+e^{i\left( k_{0}+k_{i}\right) d}\right\vert ^{2}G\left(
k_{i}\right) \delta \left( k_{i}-k_{f}\right)
\end{equation}%
and the reflection amplitude is%
\begin{equation}
\left\langle 0\right\vert a_{Lk_{f}}Sa_{Rk_{i}}^{\dagger }\left\vert
0\right\rangle =-i\gamma \left( e^{-i\phi }+e^{i\left( k_{0}+k_{i}\right)
L}\right) \left( e^{i\phi }+e^{i\left( k_{0}+k_{i}\right) d}\right) G\left(
k_{i}\right) \delta \left( k_{i}+k_{f}\right) .
\end{equation}

\section*{Supplementary Note 3: Two-photon scattering}

For two incident photons with momenta $k_{1}$ and $k_{2}$, the S-matrix reads

\begin{eqnarray}
&&\left\langle 0\right\vert A_{p_{2}+}A_{p_{1}+}SA_{k_{2}+}^{\dagger
}A_{k_{1}+}^{\dagger }\left\vert 0\right\rangle  \notag \\
&=&e^{i\left( k_{1}+k_{2}+p_{1}+p_{2}\right) T}\lim_{\left\{ \alpha
_{k}\right\} ,\left\{ \alpha _{p}\right\} \rightarrow 0}\partial _{\alpha
_{p_{2}}^{\ast }}\partial _{\alpha _{p_{1}}^{\ast }}\partial _{\alpha
_{k_{2}}}\partial _{\alpha _{k_{1}}}\mathcal{A}  \notag \\
&=&\left[ 1-i2\pi \left( \tilde{V}_{k_{2}}^{2}G\left( k_{2}\right) +\tilde{V}%
_{k_{1}}^{2}G\left( k_{1}\right) \right) \right] \delta \left(
k_{1}-p_{1}\right) \delta \left( k_{2}-p_{2}\right) +\left(
k_{1}\leftrightarrow k_{2}\right) +B,  \label{sm3}
\end{eqnarray}%
which reveals two processes: the independent scattering of the two photons
depicted by the first term in the last line of Eq. (\ref{sm3}), and the rest%
\begin{eqnarray}
B &\equiv &\int D\left[ \beta ,\beta ^{\ast }\right] \tilde{V}_{p_{2}}\tilde{%
V}_{p_{1}}\tilde{V}_{k_{2}}\tilde{V}_{k_{1}}\exp \left[ i\int_{-T}^{T}%
\mathrm{d}t\text{ }\mathcal{L}_{eff}\left( t\right) \right] \times  \notag \\
&&\int_{-T}^{T}\prod\limits_{j=1}^{4}\mathrm{d}t_{j}\text{ }\beta \left(
t_{4}\right) e^{ip_{2}t_{4}}\beta \left( t_{3}\right) e^{ip_{1}t_{3}}\beta
^{\ast }\left( t_{2}\right) e^{-ik_{2}t_{2}}\beta ^{\ast }\left(
t_{1}\right) e^{-ik_{1}t_{1}}.
\end{eqnarray}%
corresponds to the background fluorescence. Under the Fourier transform of $%
\beta \left( t\right) $ and the Dyson expansion \cite{fetter2012quantum}, $B$
becomes

\begin{eqnarray}
B &=&-\tilde{V}_{p_{2}}\tilde{V}_{p_{1}}\tilde{V}_{k_{2}}\tilde{V}%
_{k_{1}}\left( 2\pi \right) ^{2}G\left( k_{1}\right) G\left( k_{2}\right) %
\left[ \delta \left( k_{1}-p_{1}\right) \delta \left( k_{2}-p_{2}\right)
+\delta \left( k_{1}-p_{2}\right) \delta \left( k_{2}-p_{1}\right) \right]
\notag \\
&&-4i\pi \tilde{V}_{p_{2}}\tilde{V}_{p_{1}}\tilde{V}_{k_{2}}\tilde{V}%
_{k_{1}}G\left( k_{1}\right) ^{-1}G\left( k_{2}\right) ^{-1}T_{S}G\left(
p_{1}\right) ^{-1}G\left( p_{2}\right) ^{-1}\delta \left(
k_{1}+k_{2}-p_{1}-p_{2}\right) ,
\end{eqnarray}%
where the the giant system's $T$-matrix can be exactly calculated as%
\begin{equation}
T_{S}=\frac{U}{1-i\frac{U}{2\pi }\int \mathrm{d}\omega G\left( \omega
\right) G\left( k_{1}+k_{2}-\omega \right) }.
\end{equation}%
Here, the convolution of two Green's function can be calculated by expanding
the term $\cos \phi e^{i\left( k_{0}+\omega \right) L}$ in$\ G\left( \omega
\right) $ as%
\begin{eqnarray}
&&\int \mathrm{d}\omega G\left( \omega \right) G\left( k_{1}+k_{2}-\omega
\right)  \notag \\
&=&\int d\omega \sum_{mn}\frac{e^{im\left( k_{0}+\omega \right) L}\left(
-i2\gamma \cos \phi \right) ^{m}}{\left( \omega +i2\gamma \right) ^{m+1}}%
\frac{e^{in\left( k_{0}+k_{1}+k_{2}-\omega \right) L}\left( -i2\gamma \cos
\phi \right) ^{n}}{\left( k_{1}+k_{2}-\omega +i2\gamma \right) ^{n+1}}
\notag \\
&=&\frac{2\pi }{i\bar{k}_{i}-2\gamma }\sum_{mn}F_{mn}\left( \bar{k}%
_{i}\right) \Theta \left( m-n\right) .
\end{eqnarray}%
where $\bar{k}_{i}=\frac{k_{1}+k_{2}}{2}$ is the average momentum of the
incident photons, and $F_{mn}\left( k\right) $ is defined as%
\begin{equation}
F_{mn}\left( k\right) =\frac{e^{\left( m-n\right) \left( ik-2\gamma \right)
d}}{m!}\left( \frac{-i\gamma \cos \phi e^{i\theta _{Rk}}}{k+i2\gamma }%
\right) ^{m+n}\sum_{l=0}^{n}C_{mnl}\left( k\right) ,
\end{equation}%
with%
\begin{equation}
C_{mnl}\left( k\right) =\frac{\left( m+n-l\right) !}{\left( n-l\right) !l!}%
\left[ -i2\left( m-n\right) \left( k+i2\gamma \right) d\right] ^{l}.
\end{equation}%
\ With Eq. (\ref{sm4}), we can obtain the S-matrix $%
S_{p_{2}p_{1},k_{i}k_{i}} $ for two transmitted photons shown in Eq. (7).

Performing the Fourier transform to $S_{p_{2}p_{1},k_{2}k_{1}}$, the
wavefunction $w_{t}\left( X_{1},X_{2}\right) $ for two transmitted photons
at positions $X_{1}$ and $X_{2}$ is%
\begin{eqnarray}
&&w_{t}\left( X_{1},X_{2}\right)  \notag \\
&=&\frac{1}{2\pi }t\left( k_{1}\right) t\left( k_{2}\right) \left(
e^{ik_{2}X_{1}}e^{ik_{1}X_{2}}+e^{ik_{1}X_{1}}e^{ik_{2}X_{2}}\right)  \notag
\\
&&+i\frac{\gamma ^{2}}{2\pi }\left[ e^{i\phi }+e^{i\left( k_{0}+k_{1}\right)
d}\right] \left[ e^{i\phi }+e^{i\left( k_{0}+k_{2}\right) d}\right] TG\left(
k_{1}\right) G\left( k_{2}\right) F_{t}\left( X_{1},X_{2}\right)
\end{eqnarray}%
where the correlated part%
\begin{eqnarray}
&&F_{t}\left( X_{1},X_{2}\right)  \notag \\
&\equiv &\frac{1}{\pi }\int \mathrm{d}pe^{ipX_{1}}e^{i\left(
k_{1}+k_{2}-p\right) X_{2}}\left( e^{-i\phi }+e^{-i\left( k_{0}+p\right)
d}\right) \left( e^{-i\phi }+e^{-i\left( k_{0}+2\bar{k}_{i}-p\right)
d}\right) G\left( p\right) G\left( k_{1}+k_{2}-p\right)  \notag \\
&=&\frac{1}{\pi }e^{i2\bar{k}_{i}X_{2}}\sum_{mn}\left( -i2\gamma \cos \phi
\right) ^{m+n}\left( -\right) ^{n+1}e^{imk_{0}L}e^{in\left( k_{0}+2\bar{k}%
_{i}\right) L}\int dp\frac{e^{ipX_{mn}}\left( e^{-i\phi }+e^{-i\left(
k_{0}+p\right) d}\right) \left( e^{-i\phi }+e^{-i\left( k_{0}+2\bar{k}%
_{i}-p\right) d}\right) }{\left( p+i2\gamma \right) ^{m+1}\left( p-2\bar{k}%
_{i}-i2\gamma \right) ^{n+1}}  \notag \\
&=&-\frac{ie^{i\bar{k}_{i}\left( X_{1}+X_{2}\right) }}{\bar{k}_{i}+i2\gamma }%
\sum_{mn}\frac{1}{m!}\left( \frac{-i\gamma \cos \phi }{\bar{k}_{i}+i2\gamma }%
e^{i\left( k_{0}+\bar{k}_{i}\right) d}\right) ^{m+n}\{\left( e^{-i2\phi
}+e^{-i2\left( k_{0}+\bar{k}_{i}\right) d}\right) e^{\left( i\bar{k}%
_{i}-2\gamma \right) X_{mn}}\sum_{l=0}^{n}\tilde{C}_{mnl}\left( k\right)
\Theta \left( X_{mn}\right)  \notag \\
&&+e^{-i\left( k_{0}+\bar{k}_{i}\right) L-i\phi }\sum_{\alpha =\pm
}e^{\left( i\bar{k}_{i}-2\gamma \right) X_{mn}^{\alpha }}\sum_{l=0}^{n}%
\tilde{C}_{mnl}^{\alpha }\left( k\right) \Theta \left( X_{mn}^{\alpha
}\right) \}+\left( X_{1}\longleftrightarrow X_{2}\right) ,
\end{eqnarray}%
with%
\begin{equation}
\tilde{C}_{mnl}\left( k\right) =\frac{\left( m+n-l\right) !}{\left(
n-l\right) !l!}\left[ -i2\left( \bar{k}_{i}+i2\gamma \right) \left( \left(
X_{1}-X_{2}\right) +\left( m-n\right) L\right) \right] ^{l},
\end{equation}%
and%
\begin{equation}
\tilde{C}_{mnl}^{\pm }\left( k\right) =\frac{\left( m+n-l\right) !}{\left(
n-l\right) !l!}\left[ -i2\left( \bar{k}_{i}+i2\gamma \right) \left( \left(
X_{1}-X_{2}\right) +\left( m-n\pm 1\right) L\right) \right] ^{l}.
\end{equation}%
Here, the effective distances $X_{mn}$ and $X_{mn}^{\pm }$ are defined as%
\begin{equation}
X_{mn}=X_{1}-X_{2}+\left( m-n\right) L
\end{equation}%
and%
\begin{equation}
X_{mn}^{\pm }=X_{1}-X_{2}+\left( m-n\pm 1\right) L.
\end{equation}%
In the limit where $\left\vert X_{1}-X_{2}\right\vert $ goes to infinity,
the contribution to the two-photon wavefunction from the background
fluorescence vanishes, and the wave function%
\begin{equation}
w_{t}\left( X_{1},X_{2}\right) \rightarrow \frac{1}{2\pi }t\left(
k_{1}\right) t\left( k_{2}\right) \left(
e^{ik_{2}X_{1}}e^{ik_{1}X_{2}}+e^{ik_{1}X_{1}}e^{ik_{2}X_{2}}\right) .
\end{equation}

\section*{Supplementary Note 4: $N$ coupling points}

With $N$ coupling points with the same coupling strength $V$ and diffirent
phase $\phi _{n}$, the interaction between the giant cavity and the
waveguide is $H_{1}=\int_{-\infty }^{+\infty }\tilde{V}_{k}\left( b^{\dag
}A_{+k}+\mathrm{H.c.}\right) \mathrm{d}k\mathrm{,}$where%
\begin{equation}
\tilde{V}_{k}=V\sqrt{2\sum_{m,n=0}^{N}e^{i\left( \phi _{m}-\phi _{n}\right)
}\cos \left[ \left( m-n\right) \left( k_{0}+k\right) d\right] }
\end{equation}%
and the coupled mode $A_{+k}$ and free mode $A_{-k}$ are defined as
\begin{equation}
\left(
\begin{array}{c}
A_{+k} \\
A_{-k}%
\end{array}%
\right) =U_{k}\left(
\begin{array}{c}
a_{L-k} \\
a_{Rk}%
\end{array}%
\right) ,
\end{equation}%
with the unitary transformation%
\begin{equation}
U_{k}=\frac{V}{\tilde{V}_{k}}\left(
\begin{array}{cc}
\sum_{n=1}^{N}e^{i\phi _{n}}e^{-i\left( n-1\right) \left( k_{0}+k\right) d}
& \sum_{n=1}^{N}e^{i\phi _{n}}e^{i\left( n-1\right) \left( k_{0}+k\right) d}
\\
\sum_{n=1}^{N}e^{-i\phi _{n}}e^{-i\left( n-1\right) \left( k_{0}+k\right) d}
& -\sum_{n=1}^{N}e^{-i\phi _{n}}e^{i\left( n-1\right) \left( k_{0}+k\right)
d}%
\end{array}%
\right) .
\end{equation}

The self-energy is acquired as%
\begin{equation}
\Sigma \left( \omega \right) =-i\gamma \left[ N+2\sum_{m>n}^{N}\cos \left(
\phi _{m}-\phi _{n}\right) e^{i\left( m-n\right) \left( k_{0}+\omega \right)
d}\right]  \label{sm5}
\end{equation}%
where $\gamma =2\pi V^{2}$. For photon scattering, one only needs to replace
the self-energy in the $N=2$ case by $\Sigma \left( \omega \right) $ in Eq. (%
\ref{sm5}), as well as changing $\eta _{Rk}$ by
\begin{equation}
\eta _{Rk}=\sum_{n=1}^{N}e^{i\phi _{n}}e^{i\left( n-1\right) \left(
k_{0}+k\right) d}.
\end{equation}%
For example, the transmission amplitude $t\left( k_{i}\right) =\left\langle
k_{f}^{R}\right\vert S\left\vert k_{i}^{R}\right\rangle $ is $t\left(
k_{i}\right) =\delta \left( k_{i}-k_{f}\right) -i\gamma \left\vert \eta
_{Rki}\right\vert ^{2}G\left( k_{i}\right) \delta \left( k_{i}-k_{f}\right) $%
, where the Green's function $G\left( \omega \right) =\frac{1}{\omega
-\Sigma \left( \omega \right) }$. The S-matrix for two photons with
right-moving incident momenta $k_{1}$and $k_{2}$, and right moving output
momenta $p_{1}$ and $p_{2}$ is%
\begin{eqnarray}
S_{p_{2}p_{1},k_{2}k_{1}} &=&\left[ \delta \left( k_{1}-p_{1}\right) \delta
\left( k_{2}-p_{2}\right) +\delta \left( k_{1}-p_{2}\right) \delta \left(
k_{2}-p_{1}\right) \right] t\left( k_{1}\right) t\left( k_{2}\right)  \notag
\\
&&-i\frac{\gamma ^{2}}{\pi }\eta _{Rp_{2}}^{\ast }\eta _{Rp_{1}}^{\ast }\eta
_{Rk_{2}}\eta _{Rk_{1}}G\left( k_{1}\right) G\left( k_{2}\right)
T_{S}G\left( p_{1}\right) G\left( p_{2}\right) \delta \left(
k_{1}+k_{2}-p_{1}-p_{2}\right) ,
\end{eqnarray}%
where the $T_{s}$-matrix is%
\begin{equation}
T_{S}=\frac{U}{1-i\frac{U}{2\pi }\int \mathrm{d}\omega G\left( \omega
\right) G\left( k_{1}+k_{2}-\omega \right) }.
\end{equation}

The condition where the dynamics of the giant system is Markovian is
equivalent to the condition when $\tilde{V}_{k}$ independent of $k$. It gives%
\begin{equation}
\phi _{1}=\pm \frac{\pi }{2}
\end{equation}%
and%
\begin{equation}
\sum_{n=1}^{N-1}\cos \left( \phi _{n}-\phi _{n+1}\right)
=0,\sum_{n=1}^{N-2}\cos \left( \phi _{n}-\phi _{n+2}\right)
=0,\sum_{n=1}^{N-3}\cos \left( \phi _{n}-\phi _{n+3}\right) =0,...
\end{equation}%
For simplicity, we have set $\phi _{N}=0$. When these conditions can be
fulfilled, the dynamics of the giant system can be exactlydepicted by the
master equation%
\begin{equation}
\partial _{t}\rho =-i\left[ H_{S},\rho \right] +N\gamma \left( 2b\rho
b^{\dagger }-\left\{ b^{\dagger }b,\rho \right\} \right) ,
\end{equation}%
where $\gamma =2\pi V^{2}$. The input-output formalism read%
\begin{equation}
a_{out}^{l}\left( t\right) =a_{in}^{l}\left( t\right) -i\sqrt{\gamma }%
\sum_{n=1}^{N}e^{-i\phi _{n}}e^{i\left( n-1\right) k_{0}L}b\left( t-\left(
n-1\right) d\right) ,
\end{equation}%
\begin{equation}
a_{out}^{r}\left( t\right) =a_{in}^{r}\left( t\right) -i\sqrt{\gamma }%
\sum_{n=1}^{N}e^{-i\phi _{n}}e^{-i\left( n-1\right) k_{0}L}b\left( t+\left(
n-1\right) d\right) .
\end{equation}

With $M=3$, the conditions can be fulfilled with $\phi _{1}=\pm \pi /2$, $%
\phi _{2}=\mp \pi /4$, resulting in $\tilde{V}_{k}=\sqrt{6}V$. In this case,
the master equation reads%
\begin{equation}
\partial _{t}\rho =-i\left[ H_{S},\rho \right] +3\gamma \left( 2b\rho
b^{\dagger }-\left\{ b^{\dagger }b,\rho \right\} \right) ,
\end{equation}%
with the input-output formalism as%
\begin{equation}
a_{out}^{l}\left( t\right) =a_{in}^{l}\left( t\right) -i\sqrt{\gamma }\left[
e^{-i\phi _{1}}b\left( t\right) +e^{-i\phi _{2}}e^{ik_{0}d}b\left(
t-d\right) +e^{i2k_{0}L}b\left( t-2d\right) \right] ,
\end{equation}%
\begin{equation}
a_{out}^{r}\left( t\right) =a_{in}^{r}\left( t\right) -i\sqrt{\gamma }\left[
e^{-i\phi _{1}}b\left( t\right) +e^{-i\phi _{2}}e^{-ik_{0}d}b\left(
t+d\right) +e^{-i2k_{0}L}b\left( t+2d\right) \right] .
\end{equation}

We note that for $N>2$, the chiral condition is not as simple as in the $N=2$%
. In fact, for a given incident momentum $k_{i}$, one need to solve a
transcendental equation $\sum_{n=1}^{N}e^{i\phi _{n}}e^{-i\left( n-1\right)
\left( k_{0}-k\right) d}=0$ to determine the phase $\phi _{n}$.




\end{document}